\documentclass[12pt,preprint]{aastex62}

\usepackage{graphicx,epstopdf}
\usepackage{natbib}
\usepackage{bookmark}
\DeclareGraphicsExtensions{.ps}
\newcommand{\feka}{\hbox{Fe\,K$\alpha$}}

\newcommand{\msun}{\hbox{${M}_{\odot}$}}

\newcommand{\be}{\begin{equation}}
\newcommand{\ee}{\end{equation}}
\newcommand{\ba}{\begin{eqnarray}}
\newcommand{\ea}{\end{eqnarray}}

\newcommand{\chandra}{{\emph{Chandra}}}

\newcommand{\simgt}{\lower 2pt \hbox{$\, \buildrel {\scriptstyle >}\over {\scriptstyle\sim}\,$}}
\newcommand{\simlt}{\lower 2pt \hbox{$\, \buildrel {\scriptstyle <}\over {\scriptstyle\sim}\,$}}
\newcommand{\ls}{\lower 2pt \hbox{$\;\scriptscriptstyle \buildrel<\over\sim\;$}}
\newcommand{\gs}{\lower 2pt \hbox{$\;\scriptscriptstyle \buildrel>\over\sim\;$}}
\newcommand{\sarc}{$^{\prime\prime}\!\!.$}

\newcommand{\rxj}{RX\,J1131$-$1231}
\newcommand{\sdss}{SDSS\,J1004$+$4112}
\newcommand{\qj}{Q\,J0158$-$4325}

\graphicspath{{./}{figures/}}

\shortauthors{Bhatiani, Dai, \& Guerras}
\usepackage{comment}
\begin{document}

%\title{\large Constraining the sub-stellar mass objects in Extragalactic systems using Quasar Microlensing}
\title{\large Confirmation of Planet-Mass Objects in Extragalactic Systems}

\correspondingauthor{Saloni Bhatiani}
\email{salonibhatiani@ou.edu}

\author[0000-0002-9044-9383]{Saloni Bhatiani}
\affil{Homer L. Dodge Department of Physics and Astronomy,
University of Oklahoma, Norman, OK 73019, USA}

\author[0000-0001-9203-2808]{Xinyu Dai}
\affil{Homer L. Dodge Department of Physics and Astronomy,
University of Oklahoma, Norman, OK 73019, USA}
\email{xdai@ou.edu}

\author{Eduardo Guerras}
\affil{Homer L. Dodge Department of Physics and Astronomy,
University of Oklahoma, Norman, OK 73019, USA}
%% Note that the \and command from previous versions of AASTeX is now
%% depreciated in this version as it is no longer necessary. AASTeX 
%% automatically takes care of all commas and "and"s between authors names.

%% AASTeX 6.2 has the new \collaboration and \nocollaboration commands to
%% provide the collaboration status of a group of authors. These commands 
%% can be used either before or after the list of corresponding authors. The
%% argument for \collaboration is the collaboration identifier. Authors are
%% encouraged to surround collaboration identifiers with ()s. The 
%% \nocollaboration command takes no argument and exists to indicate that
%% the nearby authors are not part of surrounding collaborations.

%% Mark off the abstract in the ``abstract'' environment. 
\begin{abstract}
Quasar microlensing serves as a unique probe of discrete objects within galaxies and galaxy clusters. 
Recent advancement of the technique shows that it can constrain planet-scale objects beyond our native galaxy by studying their induced microlensing signatures, the energy shift of emission lines originated in the vicinity of the black hole of high redshift background quasars.
We employ this technique to exert effective constraints on the planet-mass object distribution within two additional lens systems, Q\,J0158$-$4325 ($z_l = 0.317$) and SDSS\,J1004+4112 ($z_l = 0.68$) using \chandra\ observations of the two gravitationally-lensed quasars. 
The observed variations of the emission line peak energy can be explained as microlensing of the FeK$\alpha$ emission region induced by planet-mass microlenses. 
To corroborate this, we perform microlensing simulations to determine the probability of a caustic transiting the source region and compare this with the observed line shift rates. Our analysis yields constraints on the sub-stellar population, with masses ranging from Moon ($10^{-8} M_{\odot}$) to Jupiter ($10^{-3} M_{\odot}$) sized bodies, within these galaxy or cluster scale structures, with total mass fractions  of $\sim 3\times10^{-4}$ and $\sim 1\times10^{-4}$ with respect to halo mass for Q\,J0158$-$4325 and SDSS\,J1004+4112, respectively. Our analysis suggests that unbound planet-mass objects are universal in galaxies, and we surmise the objects to be either free-floating planets or primordial black holes.  We present the first-ever constraints on the sub-stellar mass distribution in the intra-cluster light of a galaxy cluster. Our analysis yields the most stringent limit for primordial black holes at the mass range.
\end{abstract}

%% Keywords should appear after the \end{abstract} command. 
%% See the online documentation for the full list of available subject
%% keywords and the rules for their use.


\section{Introduction} \label{sec:intro}
%\paragraph*{}

The planet to stellar scale astronomical dark matter is also known as massive compact halo objects.  It was previously constrained to be less than 10\% of the total mass of the Milky Way \citep{alcock98}.  Recently, unbound planet-mass objects have been discovered by Galactic microlensing studies \citep{sumi11,mroz17} with a mass fraction of a few of $10^{-5}$ halo mass at the super Earth or Jupiter ranges.
In the extragalactic regime, \citet{dai18} detect planet-mass objects in the Moon to Jupiter mass range with a total mass fraction of $10^{-4}$ of a $z=0.295$ lens galaxy, RX\,J1131$-$1231.  Although tiny in the mass fraction, the detections of these objects open a new window to constrain the demographics of new types of astronomical objects, and the leading candidates are free-floating planets or primordial black holes.  
A pixel-lensing technique \citep{ingrosso09, mp18} has also been proposed to detect planet-mass objects in the nearby galaxy M31, and the technique has yielded a candidate \citep{niikura19} from recent \emph{Subaru} observations.

%\paragraph*{}
Free-floating planets (FFPs) are not gravitationally bound to particular stars and are thought to be ejected or scattered due to dynamical instabilities in the early history of star/planet formation \citep{rasio96,weidzari96} or combined with other processes such as planetary stripping in stellar clusters and post-main-sequence ejection \citep{verray12}.  It is also possible that a fraction of them form directly from gravitational collapse resembling star/brown dwarf formation down to planet scales \citep{luhman12}.
The number density of FFPs not only depends on the detailed ejection processes, but also on the planet formation models.
Primordial black holes are thought to be formed during the inflation epoch from quantum fluctuations.  Therefore, these planet mass objects can either serve as a probe of star/planet formation and scattering process or fundamental physics in the very early universe in the inflation era.
%\paragraph*{}

Extragalactic microlensing is induced by individual stars or lower mass compact objects in the lensing galaxy, and manifests in the time variable magnification of its microimages. In case of quasar microlensing \citep{cr79, Wam1,koch4}, the background quasar is first macrolensed into multiple images by a foreground lensing galaxy, and then light from the quasar is further deflected by the stars within the lens as they align with the quasar image, thus resulting in uncorrelated variability in the macroimages. Quasar microlensing serves as a special tool to study the structure of the accretion disk and spatially resolve the emission region around the supermassive black hole \citep[e.g.,][]{poo7,ang8,dai10,med11b,morg8,morg12,chen11,chen12,sluse12,gue13,odow15}. 
Since quasar microlensing is produced by the granular mass distribution within the lens, it can be used to probe the stellar content of the lensing galaxy \citep{pac86,morg8,bate11,black14}.
Furthermore, \citet{dai18} show that the technique can constrain the rogue planetary distribution, using emission immediately around the black hole in the source quasar, because the Einstein ring of planet lenses match this emission size. The Einstein radius in the source plane is given as 
\begin{equation}
R_{E}=\sqrt{\frac{4GM  D_{ls}D_{os}}{c^2 D_{ol}}},
\end{equation}
where $M$ is the deflector mass and $D_{ol}$, $D_{os}$, and $D_{ls}$ are the
angular diameter distances between the observer, lens, and
source, respectively. 
Studies of nearby Seyferts reveal relativistically broadened \feka\ lines, emanating from the innermost regions of the accretion disk \citep{fb95,rey20,vaug4,young5}, and this emission is ideal to probe the microlensing signal from planet-mass objects in the lens galaxy.
Technically, the image separation induced by planet-mass lenses are of nano-arcsecs; however, since the caustic network is produced by the combination of planet and stars, we keep using the term of microlensing in the paper.

Over the last decade, \chandra\ performed X-ray monitoring observations of several gravitationally lensed quasars to constrain the extent of the emission region around the SMBH  in the X-rays \citep{char9,char12,char17,dai10,dai19, chen12,gue17,gue18}. 
%The observations were conducted using the Advanced CCD Imaging Spectrometer (ACIS; Garmire et al.  2003) on-board the \chandra\ X-ray Observatory ( Weisskopf et al. 2002). 
Among these lensed quasars, some of the systems exhibit emission line peak variations and double line features, e.g., \rxj\ \citep{char12, char17}, SDSS\,J0924+0219 \citep{chen12}, \qj, and \sdss\ \citep{char17}. 
The line peak variation and splitting have not been observed in well studied Seyferts, such as 
MCG--6--30--15 \citep{kara14}, NGC 4151 \citep{beu17}, MCG--05--23--16 \citep{dew3}, NGC 3516 \citep{turn2}, and Mrk 766 \citep{pound3}, where the \feka\ line peak shows little variability for a single source. 
Relativistic  \feka\ lines have been predicted to peak within a range of energies $\sim$ 5--8 keV conditional upon the black hole spin, observing angle, and other factors \citep{br6} for different sources; however, for a single source, such large variations in energy up to a factor of two have not been recorded.
For \chandra\ observations of lensed quasars, these line variations are uncorrelated between the images and have been frequently detected during the course of the monitoring observations. 
%Chen et al. 2012 analyzed the initial set of observations for these lensed sources and reported that the  EWs of the \feka\ lines for these systems are significantly higher than EWs of a sample of 88 nearby Seyfert galaxies. They deduced that these large EWs are a result of the microlensing of the continuum and line emission. 
\citet{char17} performed further analysis of lensed quasars with a larger set of observations and confirmed the microlensing interpretation of the observed line shifts in the X-ray spectrum of lensed quasars. 
Because of the microlensing effect, a strip of the \feka\ emission region is magnified which, in general, has a different average energy compared to that from the whole \feka\ region, and thus introduces shift or additional line peaks \citep[e.g.,][]{pop6}.
\citet{char17} also excluded other non-microlensing interpretations, e.g., emissions arising from hot spots and patches of an inhomogeneous disk, intrinsic absorption of the continuum, occultations or radiations from the ionized accretion disk. The most well-studied source, \rxj, was observed 38 times during the period of the observations, and the team has detected \feka\ line shifts in 78 out of 152 energy spectra at a $\>$90\% confidence level, of which 21 lines are detected at $\>99$\% confidence level. These line energy variation rates were interpreted by \citet{dai18} as caustic crossing events of planet-sized deflectors within the lensing galaxy.  A stars-only model is ruled out because the predicted caustic encounter rate is too low compared to the observational evidence.

%On the other hand, the lensed AGNs produce narrow Iron K alpha lines with larger EWs as these are a result of the microlensing of a relatively smaller region (several $r_g$) of the accretion disk.  Popovic (2006) simulated the flux variation of the X-ray continumm and the reflected emission for a range of accretion disc parameters and spacetime geometries around the BH and showed that the 
%\paragraph*{}
%Description of your work, your approach
%\subsection*{content of the paper}
%\paragraph*{}

%{\color{red}
This paper performs analysis of microlensing caustic encounter rates in two additional gravitationally-lensed systems, SDSS\,J1004+4112 and Q\,J0158$-$4325, to constrain the mass fraction of the unbound planet-mass objects. 
The paper is organized as follows. In Section~\ref{sec:data}, we introduce the two systems and their \chandra\ data, including our method to calculate the line shift rates. 
In Section~\ref{sec:analysis}, we describe the microlensing analysis that involves simulating magnification maps to substantiate the observed rates. We summarize the results and conclude with discussions under Section~\ref{sec:discuss}. Throughout this paper, we adopt a standard $\Lambda$CDM cosmology with $\Omega_{m} = 0.3$, $\Omega_{\Lambda} = 0.7$ and $H_{0} = 70$~km~s$^{-1}$~Mpc$^{-1}$.

\section{Observational Data} \label{sec:data}
We focus on two gravitationally lensed systems, Q\,J0158$-$4325 and SDSS\,J1004+4112. Q\,J0158$-$4325 \citep[also\ CTQ~414;][]{maza95,morg99,morg8,faure9,chen12} is a doubly lensed quasar with $z_l=0.317$, $z_s=1.29$, and an image separation of 1\sarc22. 
The second object of interest, SDSS\,J1004+4112, is a unique quintuple and large-separation lens system \citep{inada3,og4,og10,sharon5,ota6,inada8} comprising of a massive lens galaxy cluster at a redshift of $z_l=0.68$ and a source quasar at $z_s=1.734$ with a maximum image separation of 15\arcsec.

\begin{figure}
 \includegraphics[scale=0.6]{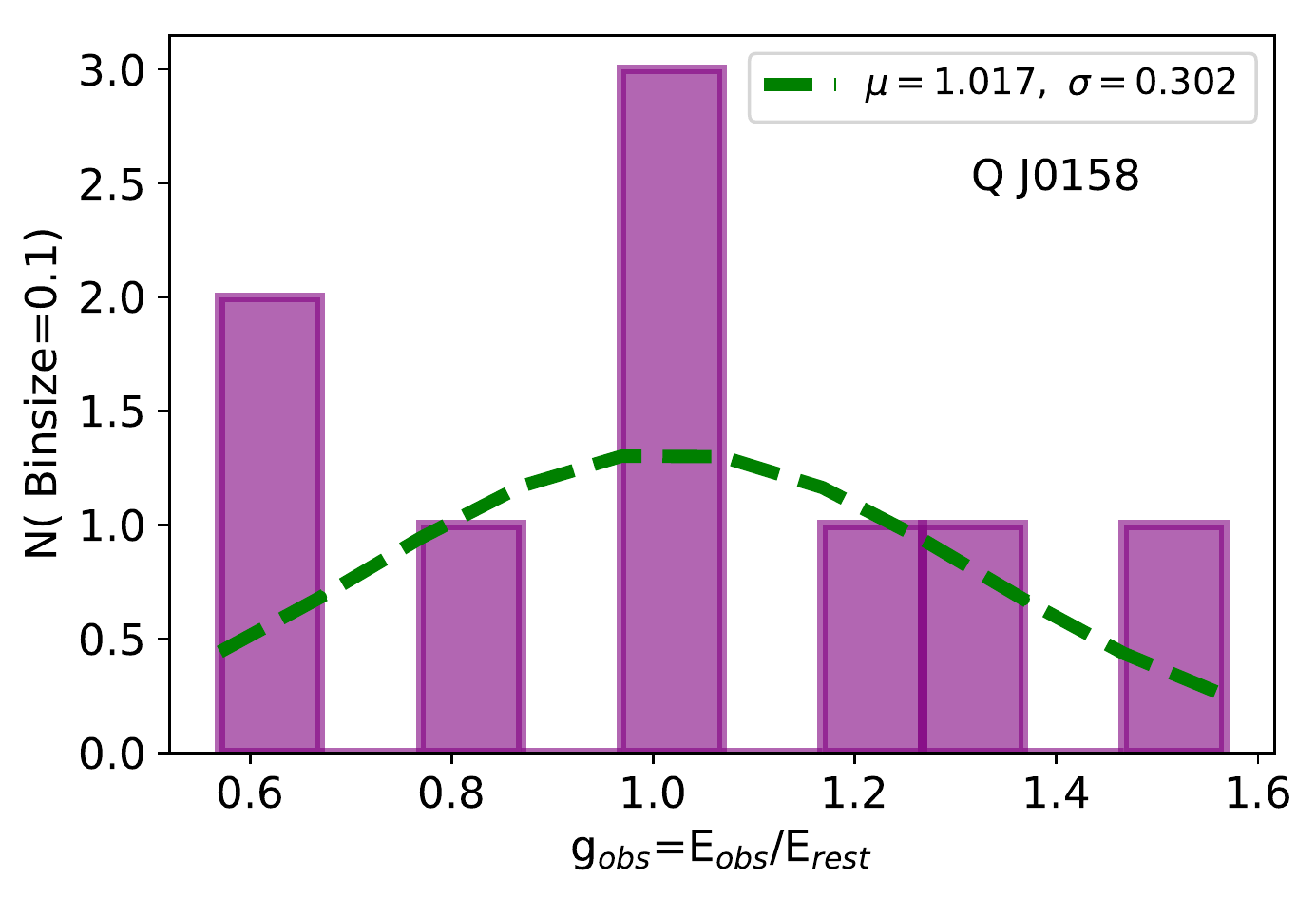}
    \includegraphics[scale=0.6]{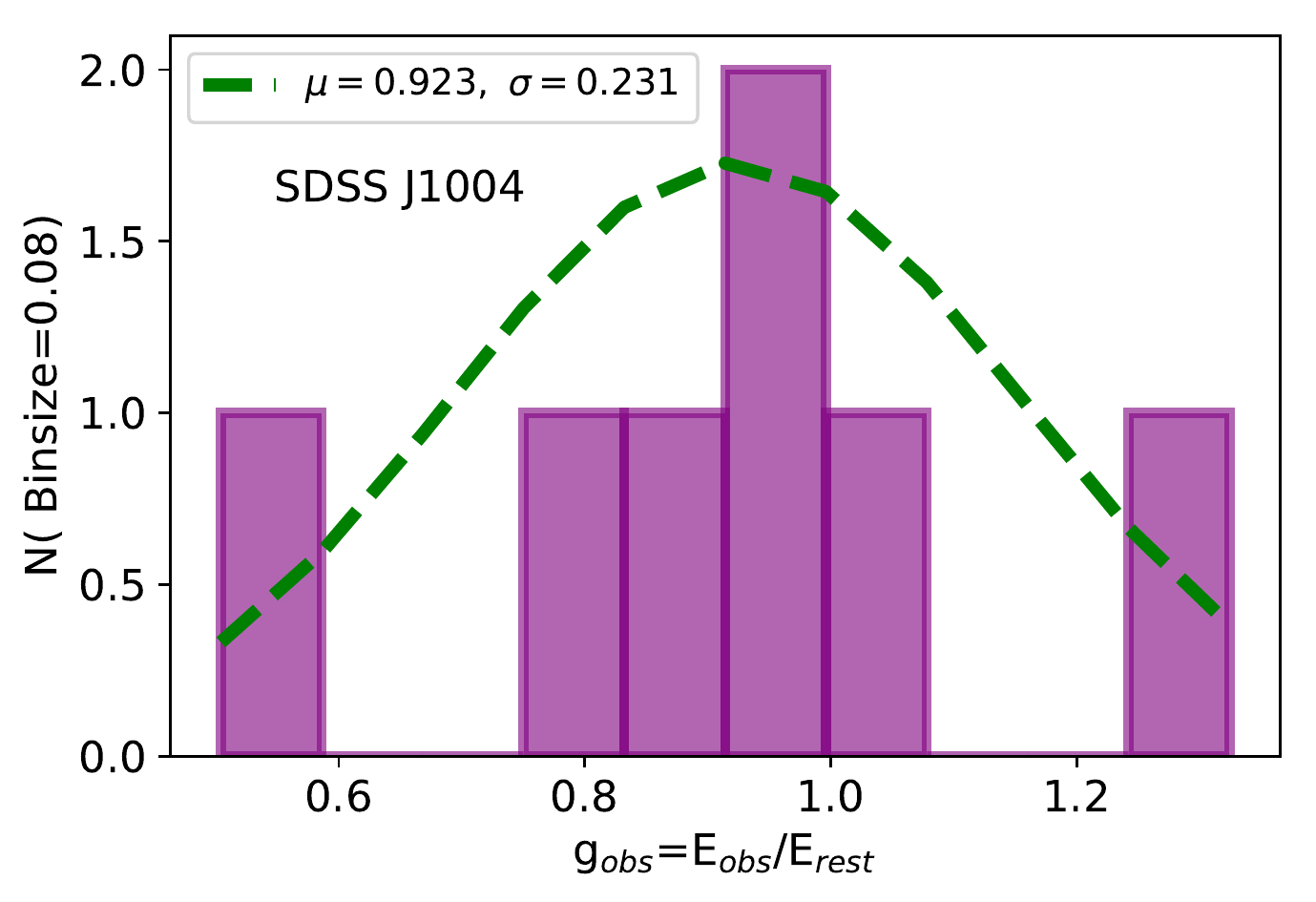}
    \caption{Distribution of the \feka\ line energy shifts for both image A and image B of \qj\ (left) and all four images of \sdss\ (right). Here we include only those lines that are detected at 90\% confidence level. A Gaussian is fit to the distribution yielding the $g_{peak}$ of the distribution. We have selected a bin size of 0.08 for \qj\ and 0.1 for \sdss\ for illustration purposes; however the choice of bin size does not significantly change the g$_{peak}$ value.
    \label{image1}}
\end{figure}

\begin{deluxetable*}{ccCrlc}[b!]
\tablecaption{Observed Line Shift Rates\label{table1}}
\tablecolumns{6}
\tablenum{1}
\tablewidth{0pt}
\tablehead{
 \colhead{Source } & 
\colhead{\chandra} & 
\multicolumn{4}{c}{Observed Line Shift Rates} \\
%\colhead{epoch of 3$\sigma$ line shift} & \colhead{Observed event rate} \\
\colhead{} &
\colhead{Pointings} &
\multicolumn{2}{c}{$>$90\% detected} &
\multicolumn{2}{c}{$>$99\% detected} \\ 
\colhead{} &
\colhead{} &
\colhead{$>1\sigma$ shift} &
\colhead{$>3\sigma$ shift} &
\colhead{$>1\sigma$ shift} &
\colhead{$>3\sigma$ shift} 
}
\startdata
\qj  \tablenotemark{a} & 12 & 0.250 & 0.208 & 0.083 & 0.083\\
A & 12 & 0.416   & 0.333   & 0.166 & 0.166 \\
B & 12 & 0.083 & 0.083 & 0.000 & 0.000 \\
\sdss \tablenotemark{b} & 10 & 0.150 & 0.125 & 0.100 & 0.075\\
A & 10 & 0.100 & 0.100 & 0.100 &  0.100\\
B & 10 & 0.200 & 0.200 & 0.000   &  0.000\\
C & 10 & 0.100 & 0.100 & 0.100 &  0.100\\
D & 10 & 0.200 & 0.100 & 0.200 &  0.100\\
\enddata
%tablecomments{The line shift rates for \qj\ and \sdss\ are calculated under different schemes. }
%\multicolumn{6}{l}{\textsuperscript{*}\footnotesize{The footnote}}
\tablenotetext{a}{ Line shift rates for the total of all images(A and B) of \qj  }
\tablenotetext{b}{ Line shift rates for the total of all images(A, B, C and D) of \sdss}
\end{deluxetable*}

These two sources have been monitored relatively frequently by \chandra, 12 times between November 2010 and June 2015 for \qj\ and 10 times between January 2005 and June 2014 for \sdss.
The observations have shown credible evidence of large redshifted or blueshifted \feka\ lines with respect to the rest frame peak energy, $E_{rest}=6.4$~keV for the two systems \citep{char17}, and moreover, double \feka\ lines have also been detected on some occasions.
Hereafter, we use the generalized Doppler shift, 
\begin{equation}
g=\frac{E_{obs}}{E_{rest}},
\end{equation}
to characterize the shifts.
Figure~\ref{image1} shows the $g$ distribution of \feka\ lines detected in \sdss\ and \qj\ \citep{char17}, ranging  between 0.6 to 1.5 for \qj\ and 0.5 to 1.3 for \sdss. 
For the 12 \chandra\ pointings of \qj, ten relativistic \feka\ lines were detected at more than 90\% confidence, among which three \feka\ lines are detected at more than 99\% confidence with one double line detection. For the ten \chandra\ pointings of \sdss, eight relativistic \feka\ lines are detected at more than 90\% confidence, and among these six \feka\ lines are detected at more than 99\% confidence with one double line detection.  
Thus, \sdss\ and \qj\ are the best targets to model the line shift rates induced by microlensing after the well-monitored lens \rxj.
These line energy shifts can be ascribed to a differential magnification of a section of the emission region bearing a range of $g$ values as a caustic transits the innermost disk, thereby resulting in the variation of observed \feka\ line profile \citep{pop3,pop6,char17,kc17,led18}. 
%This is in contrast to the non-lensed Seyfert galaxies such as MCG−6−30−15, NGC 4151, and MCG−05−23−16 for which the peaks of the \feka lines are found to be constant\citep{kar14,beu17,wang17}. 
%Thus, each event of line variation is recorded as a microlensing event and the line shifts rates of the system are measured from its $g$ distribution. Since the monitoring epochs for \sdss\ and \qj\ are sparse, we have determined the g distribution by combining the  observations for all images with line detections at a 90\% confidence level. In the case of double line detections for the same image, we select the stronger line for the purpose of our calculations. 
To calculate the line shift rate, we first identify the peak ($g_{peak}$) of the $g$ distribution by a fitting a Gaussian to the distribution, yielding $g_{peak} =0.923$ for \sdss\ and $g_{peak} =1.017 $ for \qj\ (Figure~\ref{image1}). For this purpose, we use the lines detected at $>$ 90\% confidence level (See Table 6 and Table 7 of \citet{char17}).
The significance of the line energy shift from the peak is determined for each detected lines  using their respective measured energies and uncertainties.
We list the line shift rates under different selection cuts in Table~\ref{table1}.
For the subsequent microlensing event rate analysis, we have conservatively considered only those lines that were detected at a confidence level of $>$ 99\% and exhibit a line shift of $> 3\sigma$ from the $g$ peak. 
%For both \sdss\ and \qj, we have restricted our analysis to image A alone, because the brightest image is less sensitive to selection biases.
For both \sdss\ and \qj, we have conducted an independent analysis of the brightest image A  together with a combined analysis of all the images. 
Owing to the limited S/N of our observations, we have restricted the analysis of individual images to the highest S/N image A.
\section{Microlensing Analysis} \label{sec:analysis}
To explain the occurrence of the energy shifts of the \feka\ lines emitted from the accretion disk, we perform a microlensing analysis that simulates the caustic encounter rates that contribute to the aforementioned line shifts. In quasar microlensing, the gravitational field of the local distribution of the microlenses affects the amplification of the light passing through the region of the lens galaxy. These amplification variations as a function of the source position are represented by a microlensing magnification map \citep{key86,wam90a}, comprising of a complex mesh of caustics, along which the magnification diverges, surrounded by low magnification regions. 
The statistical attributes of the maps, such as the density of caustics, are governed by three main model parameters namely $\kappa$ (convergence), $\gamma$ (shear), and $\kappa_s$ or $\kappa_*$ (surface mass density in smoothly distributed or discrete matter, respectively), where $\kappa = \kappa_s + \kappa_*$.  We also define $\alpha$ as the mass fraction of different components with respect to the total surface mass density, e.g., for the stellar component $\alpha_* = \kappa_* / \kappa$.

\begin{deluxetable*}{ccCrlc}[b!]
\tablecaption{Macro Lens model Parameters\label{table2}}
\tablecolumns{6}
\tablenum{2}
\tablewidth{5pt}
\tablehead{
\colhead{ Object} &
\colhead{ Image} &
\colhead{$\kappa$} &  %\tablenotemark{a}
\colhead{$\gamma$} & 
\colhead{$\kappa_{*}$} & 
\colhead{$\kappa_{*}$/$\kappa$}  
}
\startdata
\qj & A & 0.727 & 0.187 & 0.020 &  0.028  \\
    & B & 0.994    &    0.285  & 0.070 & 0.070  \\
\sdss & A & 0.763  & 0.300 & 0.007& 0.010 \\
 & B & 0.696  & 0.204 & 0.006& 0.010 \\
 & C & 0.635  & 0.218 & 0.006& 0.010 \\
 & D & 0.943  & 0.421 & 0.009& 0.010 \\
\enddata
%\tablenotetext{a}{At exposure start.}
%\tablecomments{ For all images of \qj\ and \sdss, local convergence, total shear, surface density of stars and stellar surface density fractions are listed.  }
\end{deluxetable*}

\begin{figure}
    \includegraphics[scale=0.60]{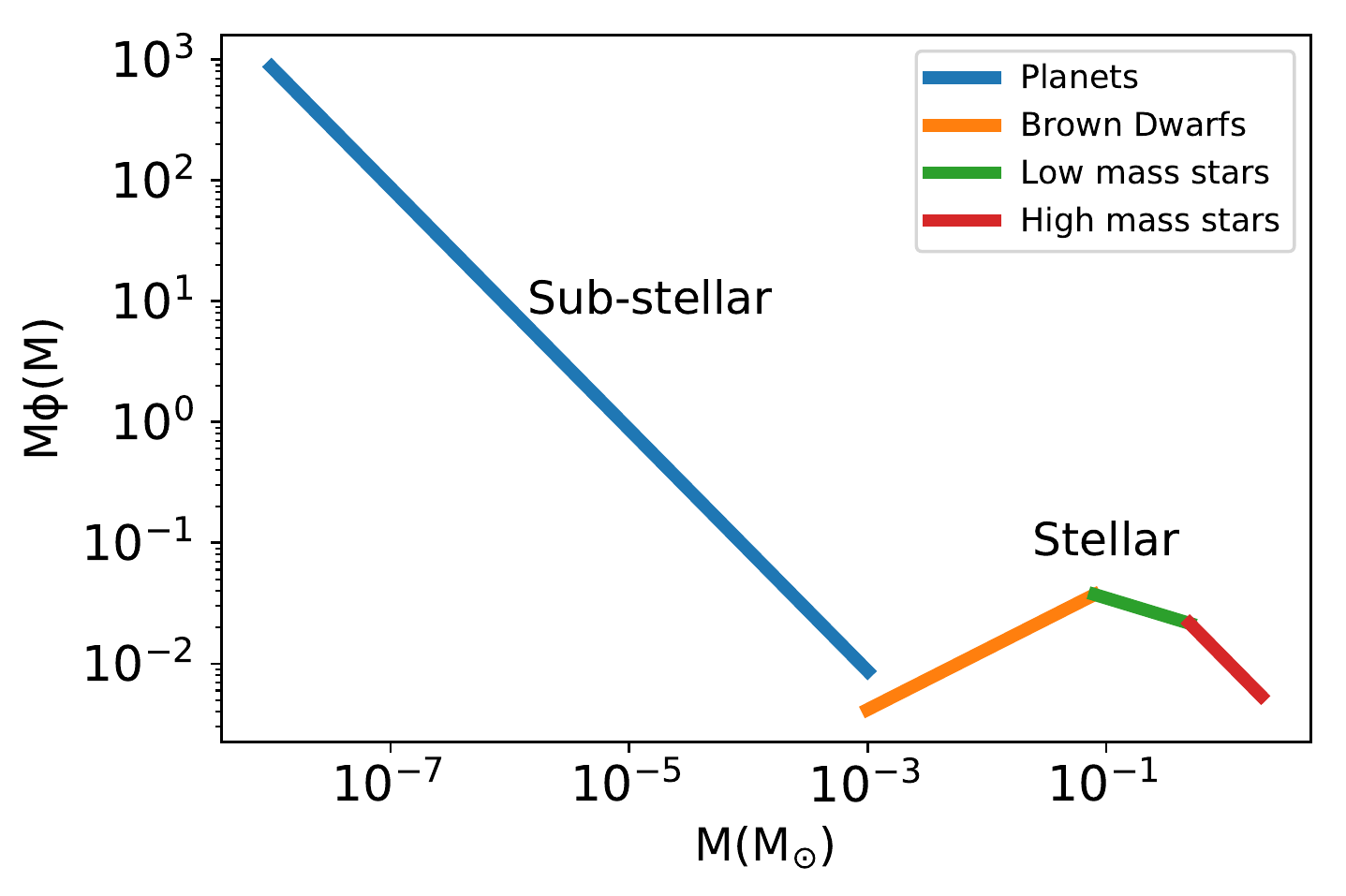}
    \includegraphics[scale=0.60]{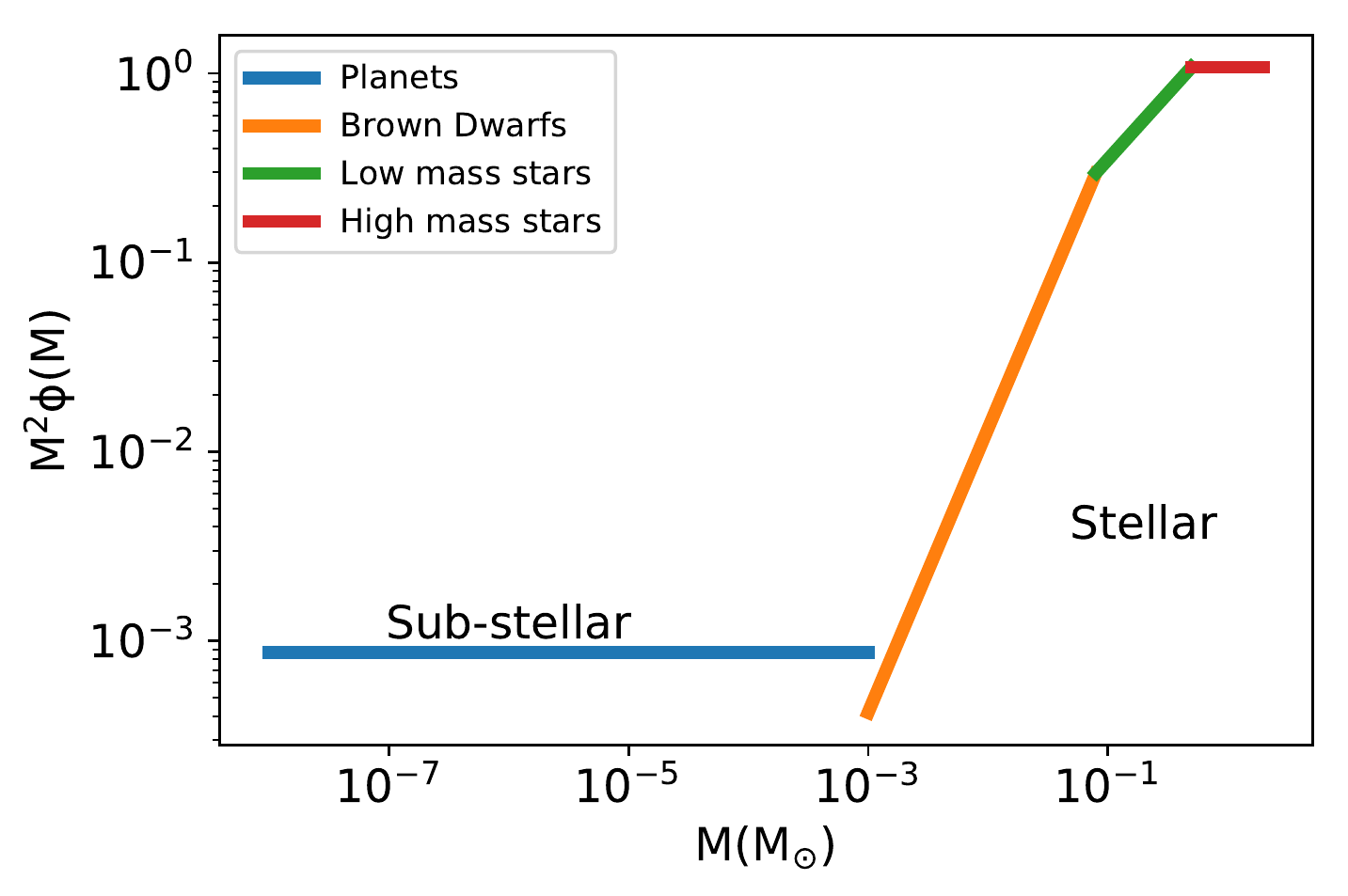}
    \begin{center}
    \includegraphics[scale=0.60]{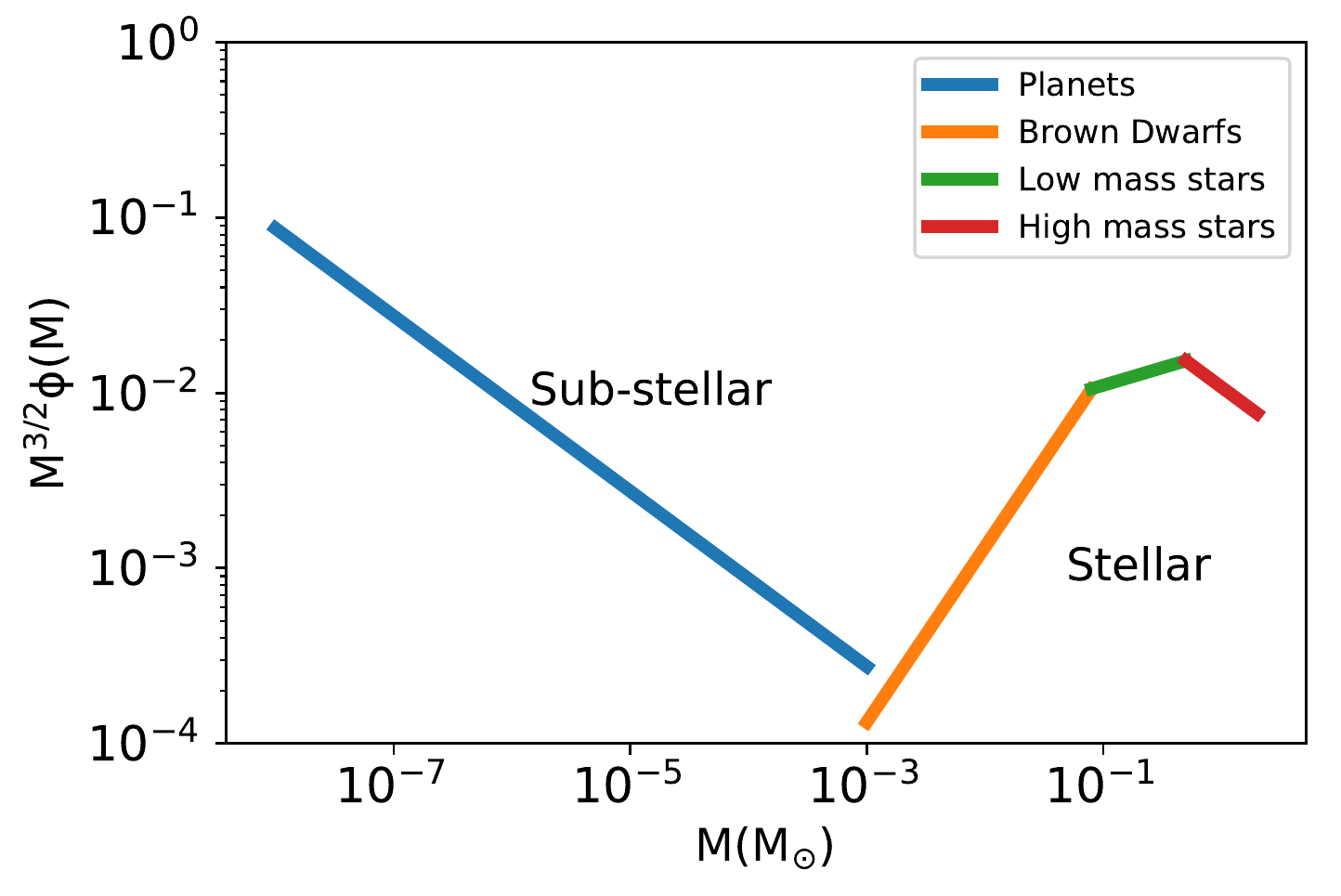}
    \end{center}
    \caption{Discrete lens mass function is plotted against the lens mass on a logarithmic scale for the stellar and planet populations with the mass fractions of $\alpha_* = 0.1$ and $\alpha_{p} = 10^{-4}$, respectively. 
    %The top-left one plots $M\phi(M)$, the top-right $M^2 \phi (M)$, and bottom-center $M^{3/2}\phi(M)$. 
    Integration of $M\phi(M)$(top-left) and $M^2 \phi (M)$(top-right) over the logarithmic mass range represents the number of microlenses and mass within the two population regimes respectively. We see that the number of microlenses is dominated by planetary objects while the total mass is dominated by stars.  The caustic density, proportional to $M^{1/2}$, is represented by the integration of $M^{3/2}\phi(M)$( bottom-center) over the logarithmic mass range. It is evident that the addition of planetary microlenses results in an increased contribution to the caustic density in contrast to the exclusively stellar scenario. 
    \label{image2}}
\end{figure}

 %$M\phi(M)$ is plotted as a function of M on a logarithmic scale for the sub-stellar population with sand the stellar population considered. The area under the curve gives the number count of objects within each population range. (Top right) Here we plot $M^{3/2} \phi(M) $as a function of M such that the area under the curve represents the microlensing effects induced by each population. (Bottom center) M$^2 \phi(M)$ is plotted as a function of M such that the area under the curve gives the mass within each population range. 

%(Left) MdN/dM is plotted as a function of M on a logarithmic scale that represents the number count of objects within each population range. We see a dominant contribution from the small sized planetary objects in contrast to the massive stellar objects. (Center ) M3/2dN/dM is plotted as a function of M(Msun) representing the microlensing effects that scale as M1/2 for both the populations. The planet sized objects show relatively higher contribution with respect to the stellar microlenses due to the contribution of edge effects. (Right) M2dN/dM is plotted as a function of M on a logarithmic scale that represents the mass within each population range. Here we see a dominant contribution to the mass from stellar sources relative to the planet sized sources.

We produce magnification maps using the inverse polygon mapping (IPM) method developed by \citet{med6, med11}.
 The macro model parameters for \qj\ are adopted from \citet{morg12}, where the lens galaxy is modeled by a combination of de Vaucouleurs and NFW components with  %\citep{dai9,morg12} using GRAVLENS\citep{keet1}. Here we produce a progression of models parametrized by f$_{ML}$, the fractional mass of the stellar component( or de Vaucouleur component) with respect to a constant M/L ratio model(f$_{ML}$=1), varying between a 10$\%$ to a 100 $\%$(no halo case) with an equal stepsize. %For each model we generated 10 realizations.%  
 %Previous time delay measurements for this system back a 
 the best fit mass-to-light ratio to be 0.1. 
% This model yields $\kappa = $, $\gamma = $ and $\kappa_{*}/\kappa = $. 
 For \sdss, we use the microlensing parameters ($\kappa$, $\gamma$) in \citet{gue17}, where they employ a cluster mass model for the lens from \citet{og10}. As a cluster lens, microlensing in \sdss\ is induced by stars and planets in the intra cluster light (ICL). We extrapolate the surface brightness of ICL to the image locations and obtain the mass fraction of $\alpha_* \sim$ 0.01. Our adopted value is aligned with the estimates of $\alpha_*$ for another galaxy cluster, MACS J1149+2223 (z$_l$ = 0.54), at similar distances to the cluster center \citep{venu17, og18}.  Table~\ref{table2} lists the global convergence, shear, stellar surface mass density, stellar mass fractions for all images of \qj\ and \sdss. 
 
\begin{figure}
    \includegraphics[scale=0.47]{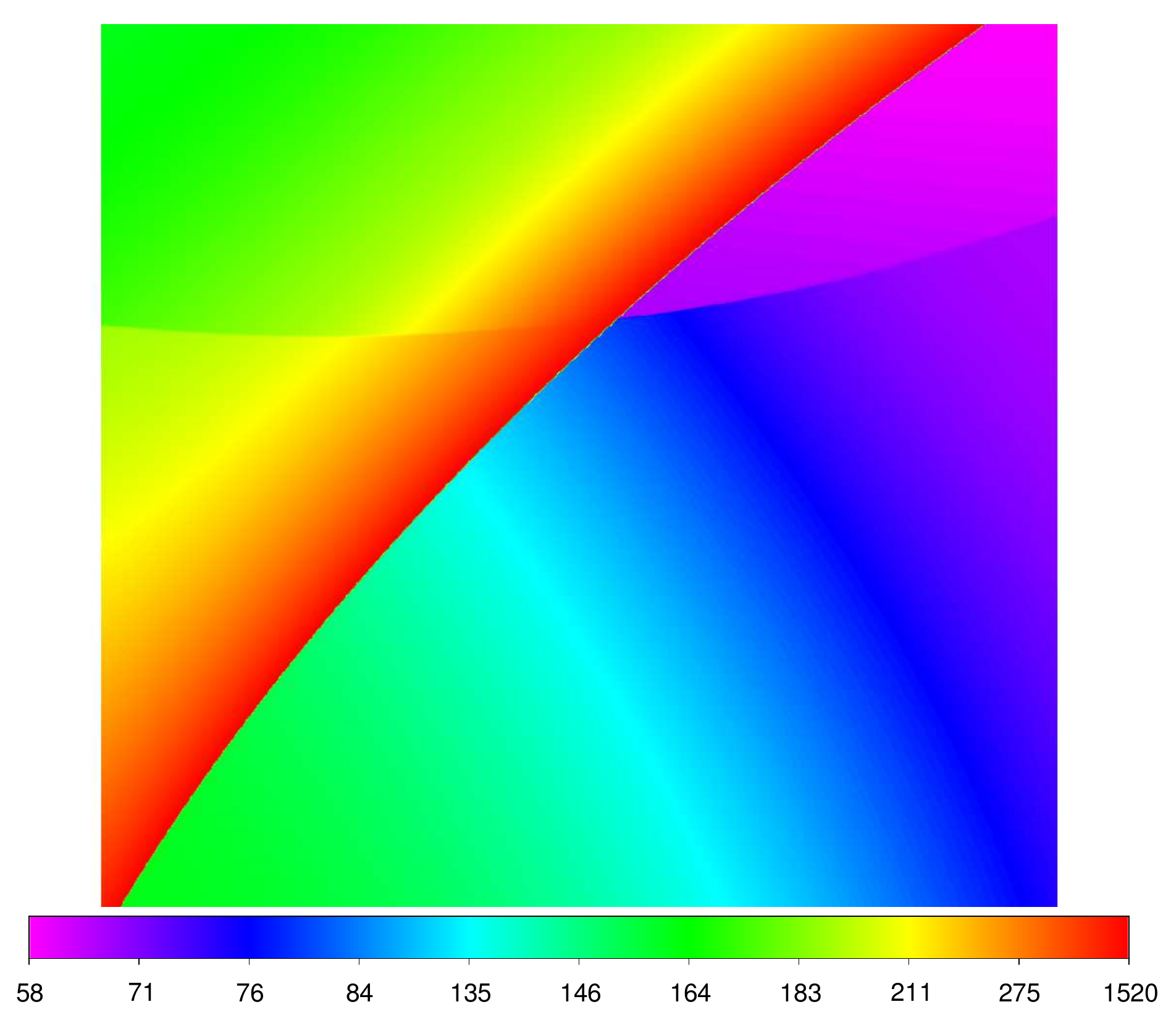}
    \includegraphics[scale=0.47]{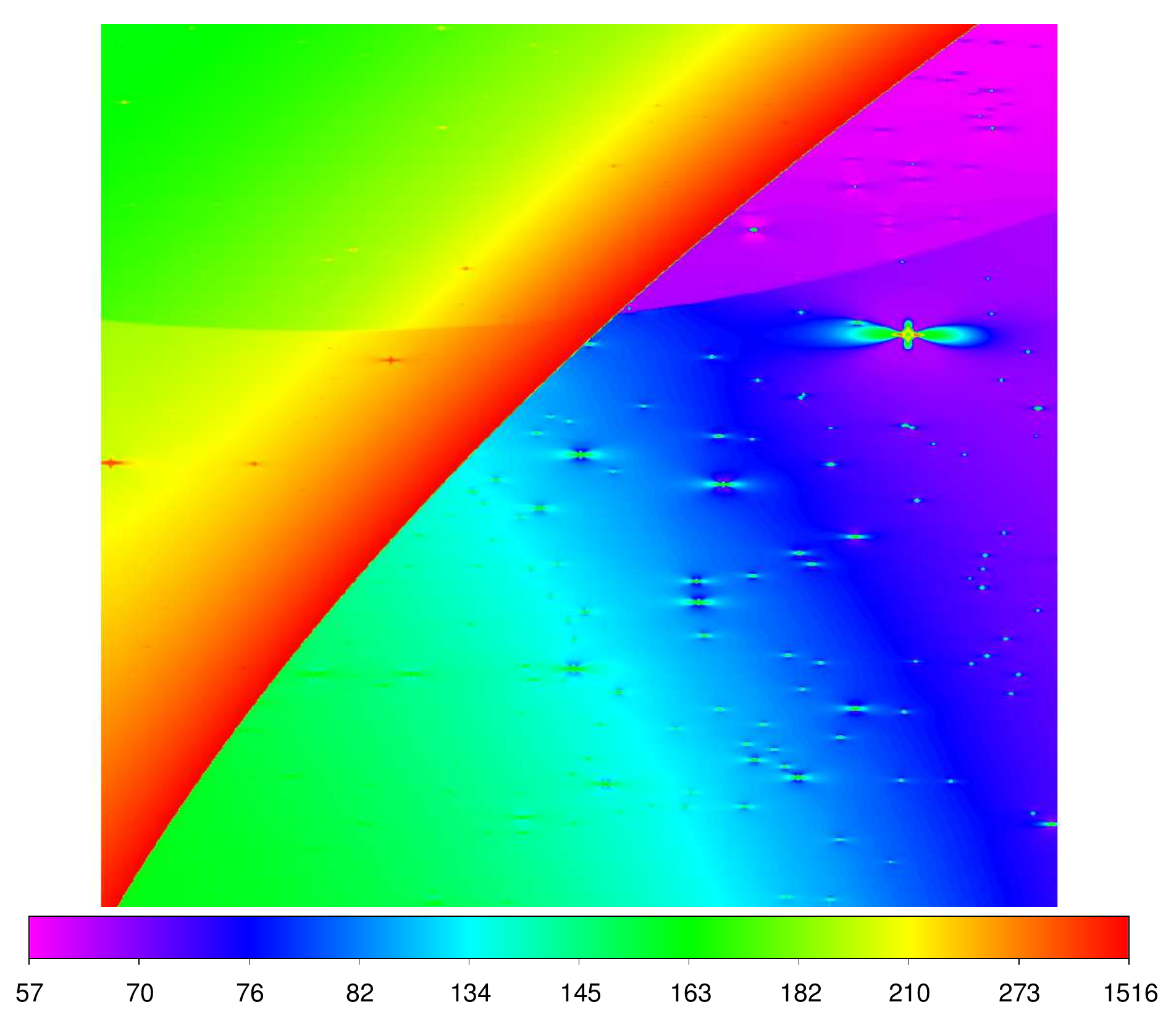}
    \includegraphics[scale=0.47]{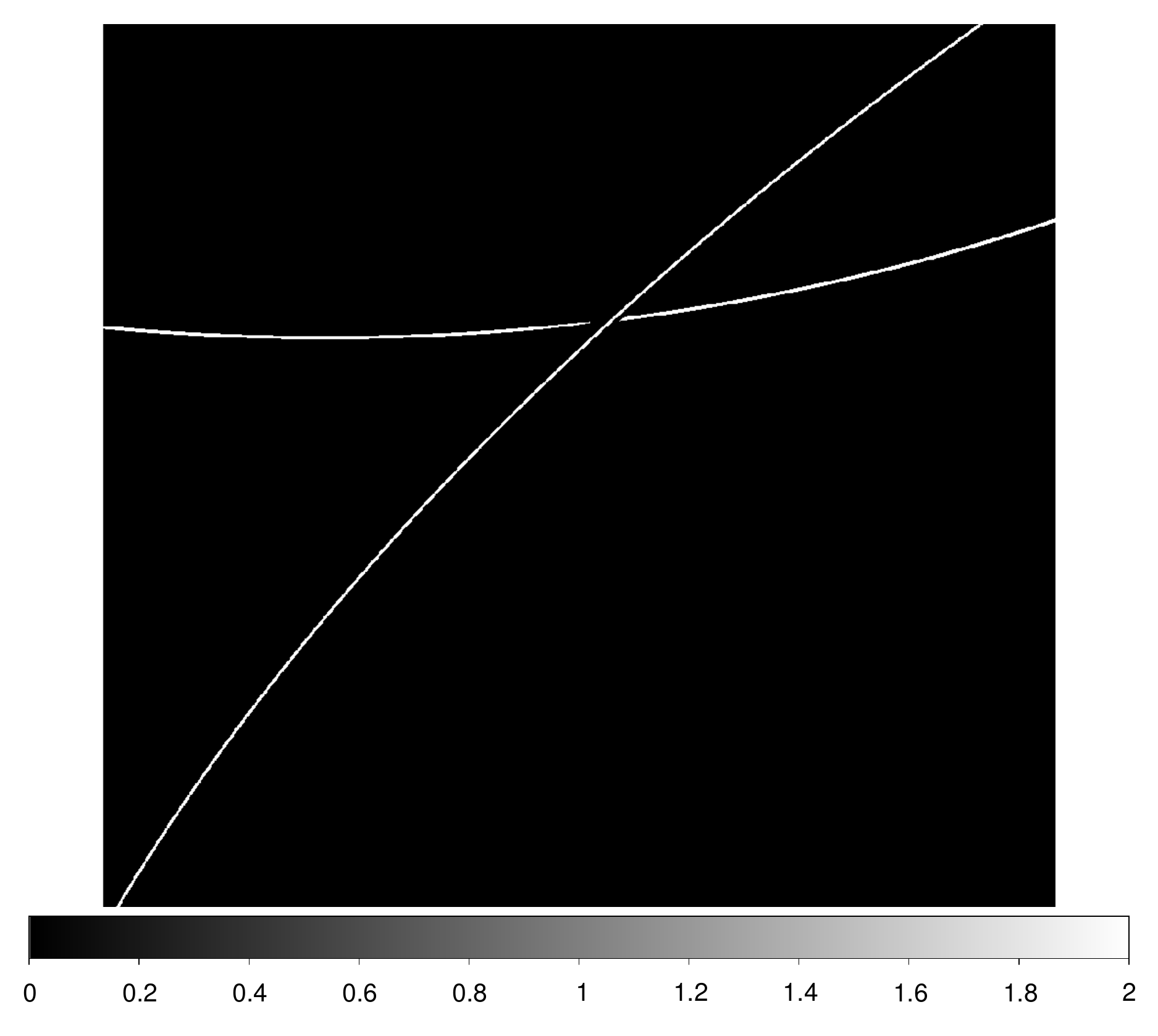}
    \includegraphics[scale=0.47]{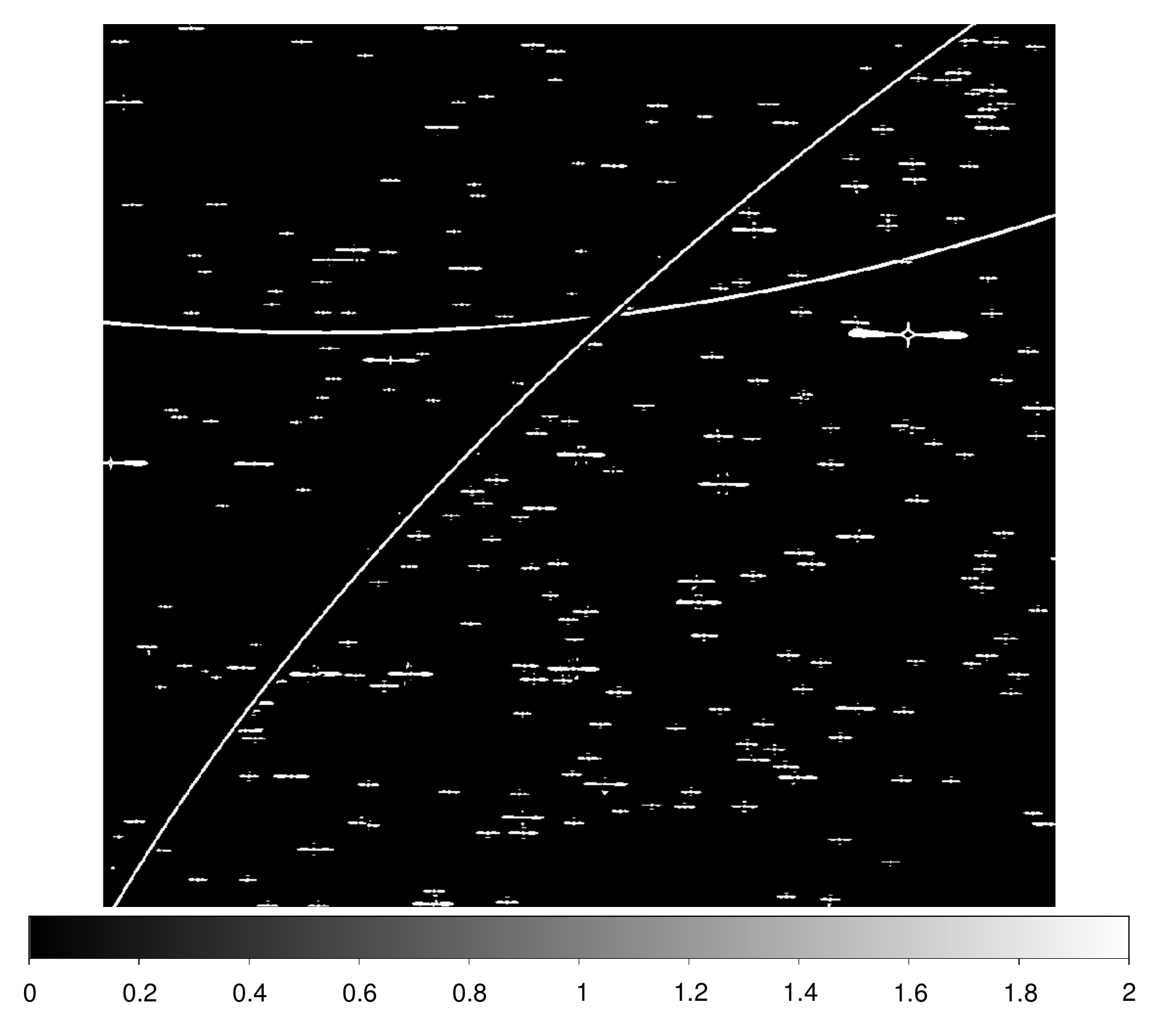}
    \caption{(Top Left) Microlensing magnification map of \qj\ with stars only and with a dimension of (400 $r_g$)$^2$. The color bars indicate the relative magnification value of the map. (Top right) The same magnification map but with the additional planet population with a planet mass fraction of $\alpha_{p} = 0.003$. The caustic density is much higher with the additional planets. (Bottom) The magnification maps convolved with a sharpening kernel with a source size of two pixels for stars only (Bottom left) and with planets (Bottom Right). We can see that the sharpened images have captured the caustic network of the map.}
    \label{image3}
\end{figure}

The  mass function for the stellar population is modeled as a three-segment power-law function, $\phi(M)\propto M^{-\eta}$, with the mass range, 0.001--0.08 \msun, corresponding to the brown dwarf population, 0.08--0.5 \msun\ for the low mass range, and 0.5--2 \msun\ for the high mass range (Figure~\ref{image2}).  The power law indices are assumed as 0.5, 1.3, and 2, respectively, for the brown dwarfs, low mass, and high mass stars \citep{dai18}. The normalization of the stellar mass function is set such that we achieve the expected stellar surface mass fractions for each image.

Subsequently, we model the sub-stellar population that includes planet-mass objects that serve as additional microlenses with a surface density of $\kappa_p$, (Figure~\ref{image2}).  We particularly focus on rogue planets, or free-floating planets \citep{dai18,sumi11,strigari12}, as the planets bound to stars are positioned way inside the Einstein ring of the parent star which is unable to produce any significant effect on the magnification maps.  
The planetary mass distribution is drawn from a power law model, $\phi_p(M)\propto M^{-\eta_p}$ with a power index of $\eta_p=2$ \citep{strigari12,dai18}. 
Here, the normalization, $\alpha_{p}=\kappa_{p}/\kappa$, is considered as a free parameter, and we have explored multiple variations to exert effective constraints. 
Hence, we generate multiple sets of magnification maps with $\alpha_{p}$ ranging from $10^{-5}$ to $10^{-3}$, corresponding to $10^2$ to $10^4$ Moon ($10^{-8}$\msun) to Jupiter ($10^{-3}$\msun) mass planets per main sequence star. 
%The upper bound considered here represents the latest limits on the floating Jovial planets in the Milky way galaxy \citep{mroz17}.  
Due to the high computation cost associated with generating extensive magnification maps that accommodate a large number of lenses, the size of the magnification maps is restricted to 400$\times$400 $r_g$ with a pixel scale of 0.375 $r_g$ (gravitational radius units). 
The dimension of the discrete lens population used in IPM is much larger to accurately calculate the magnification map. 
%It should be noted the lens plane area for star-only magnification maps is considered to be about four times larger than the lens plane area for the combined maps (including stars and planets) to effectively account for the impact of a discrete distribution of stars on a smaller source region. 
For instance, the computed magnification map for \sdss\ is of dimension 0.53 $R_E$ (Einstein radius of 1 \msun\ units), while the dimension of the deflectors used in IPM is 58.89 $R_E$ for star-only maps, while the combined map (consisting of stars and planets both) uses a square discrete lens region of the dimension 16.83 $R_E$ in the lens plane.
%If the size of a map is small compared to the typical size of a caustic (∼REin), for fixed resolution, then a particular caustic structure may dominate the map and the corresponding MPD (see fig. 1 in Wambsganss et al. 1990, and panels 61–64 of fig. 1 in W92). Therefore, varying the microlens positions on smaller maps is expected to have a larger effect.
For each set of parameters ($\kappa$, $\gamma$, $\alpha_*$, $\alpha_p$), we generate 30 random realizations of  magnification maps to sample the large scale variations in the caustic density. 
Figure~\ref{image3} shows some example magnification maps.

Following the computation of the magnification maps, we measure the caustic encounter rates as a function of the source size. 
For this, we employ an edge detection algorithm to extract the discontinuities in the maps
\citep[for an analytic approach, see also][]{witt90}.
First, we smooth the image with a small constant kernel of 3$\times$3 pixels (about 1 $ r_g^2$) to reduce the computational noise by blurring the image slightly. 
%We use a 3\times3 kernel with a constant pixel value of 1/9.
Next, we convolve the map with an $(n_s+4)\times(n_s+4)$ sharpening kernel for a source size of $n_s$ in pixels, such that the central $n_s\times n_s$ pixels bear a positive value $x$ while the rest of the background pixels carry a value of $-1$. Here $x$ is posited as $((n_s+4)^2 - n_s^2)/n_s^2$ such that when we apply this convolution matrix to the map, the pixels for which the ratio of summation over the source pixels to the summation over background pixels ($\Sigma_{s}/\Sigma_{b}$) $>$ 1 result in a positive value, whereas the remaining pixels return a negative or zero value. The source size in pixels is varied on a logarithmic scale up to $\sim 10\ r_g$. Because for a source larger than a pixel, different parts of the source experience different magnification, therefore the pliable source size, $n_s$, incorporates the finite source size effects. 
%maybe logarithmic scheme is adopted
The sharpened magnification maps vividly reveals its salient features, i.e., the caustic structure, by improving the contrast (Figure~\ref{image3}).  We determine the probability of the caustic encountering the source region by calculating the ratio of the positive pixels over all the valid pixels. For all the 30 maps attached to a given set of model parameters, the model caustic encounter probability for a particular source size is estimated as the average over all the 30 probabilities and the uncertainty is characterized by the variance of these probabilities. % Several studies 

\begin{figure}
    \includegraphics[width=0.5\textwidth]{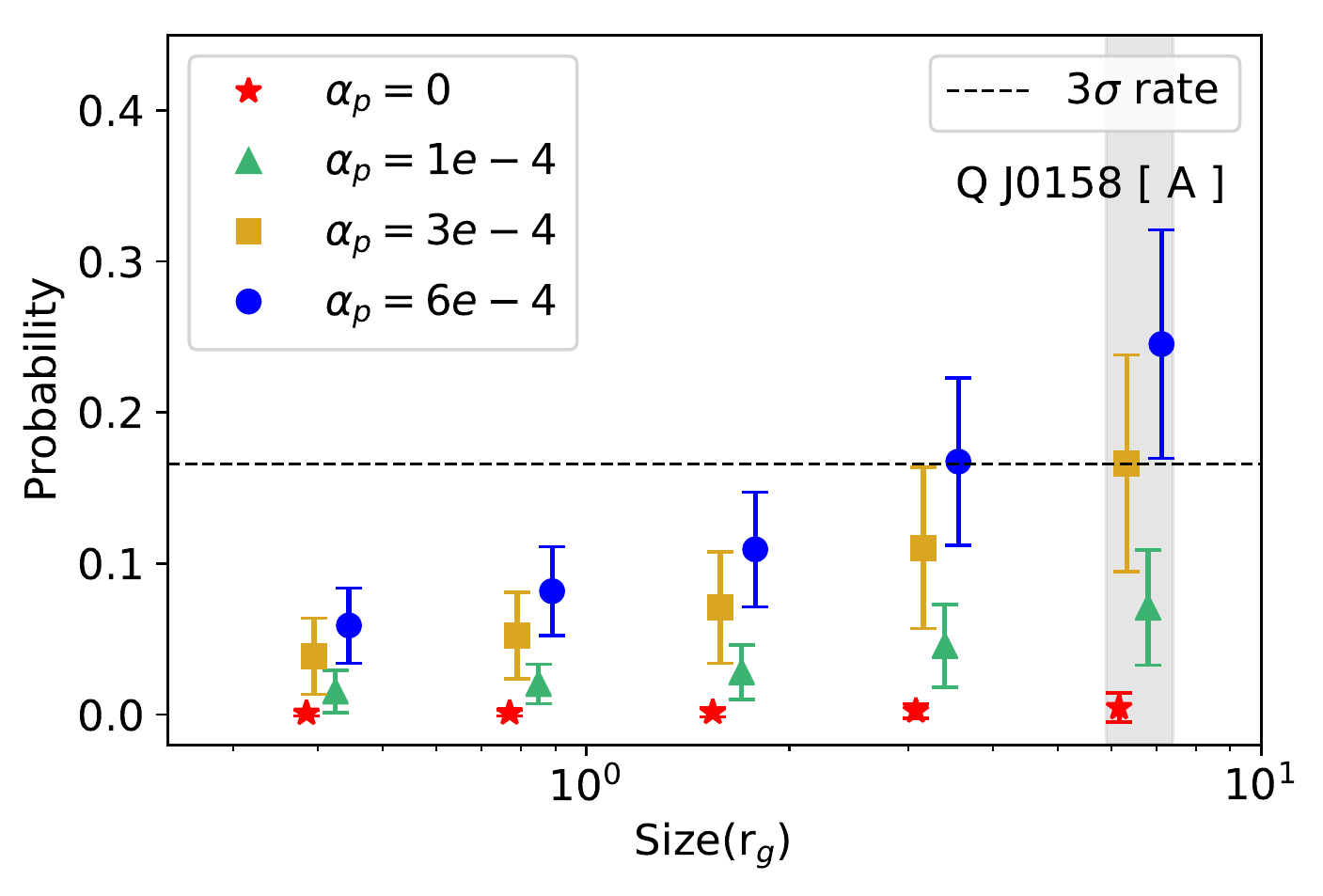}
    \includegraphics[width=0.5\textwidth]{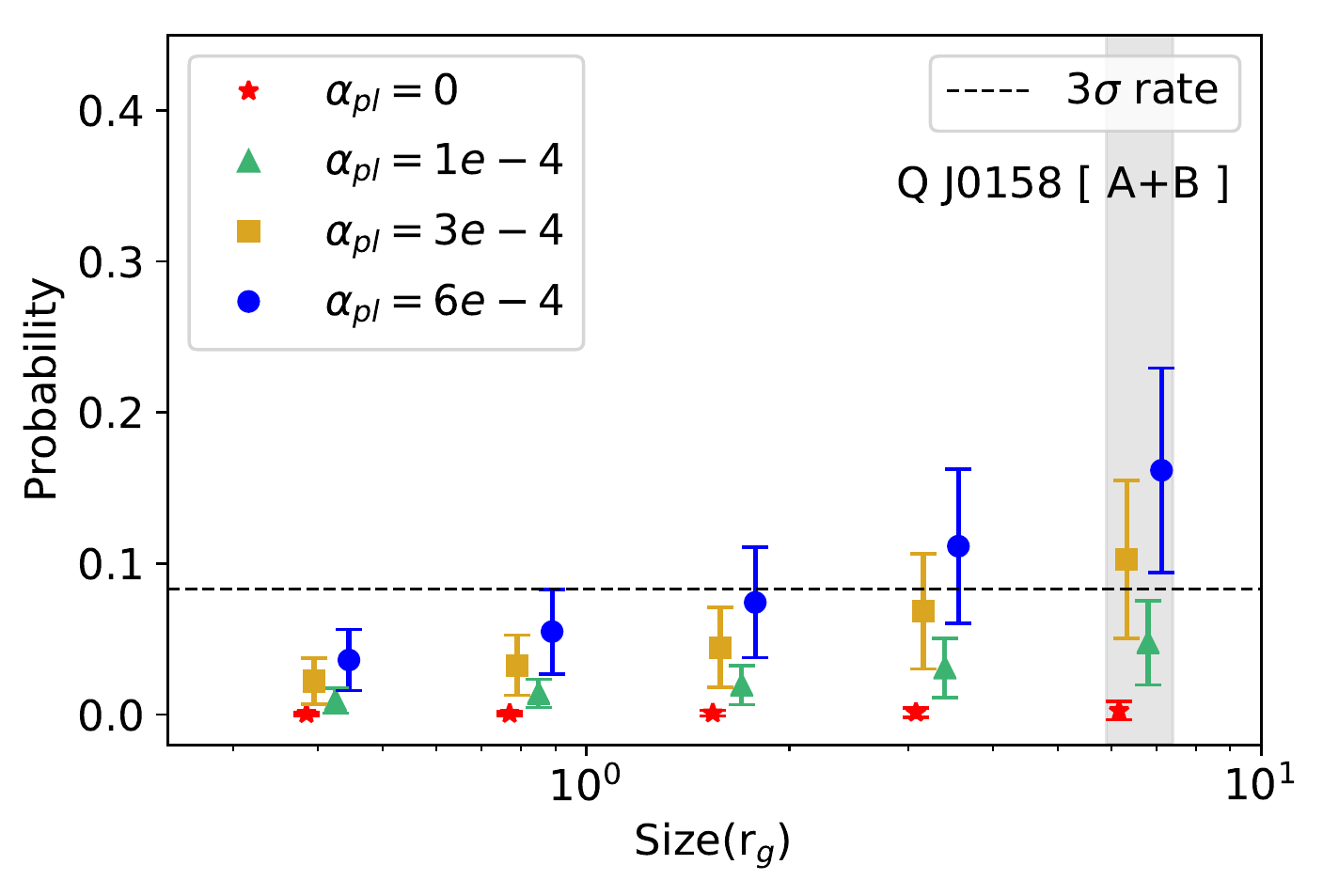}
    \\
    \includegraphics[width=0.5\textwidth]{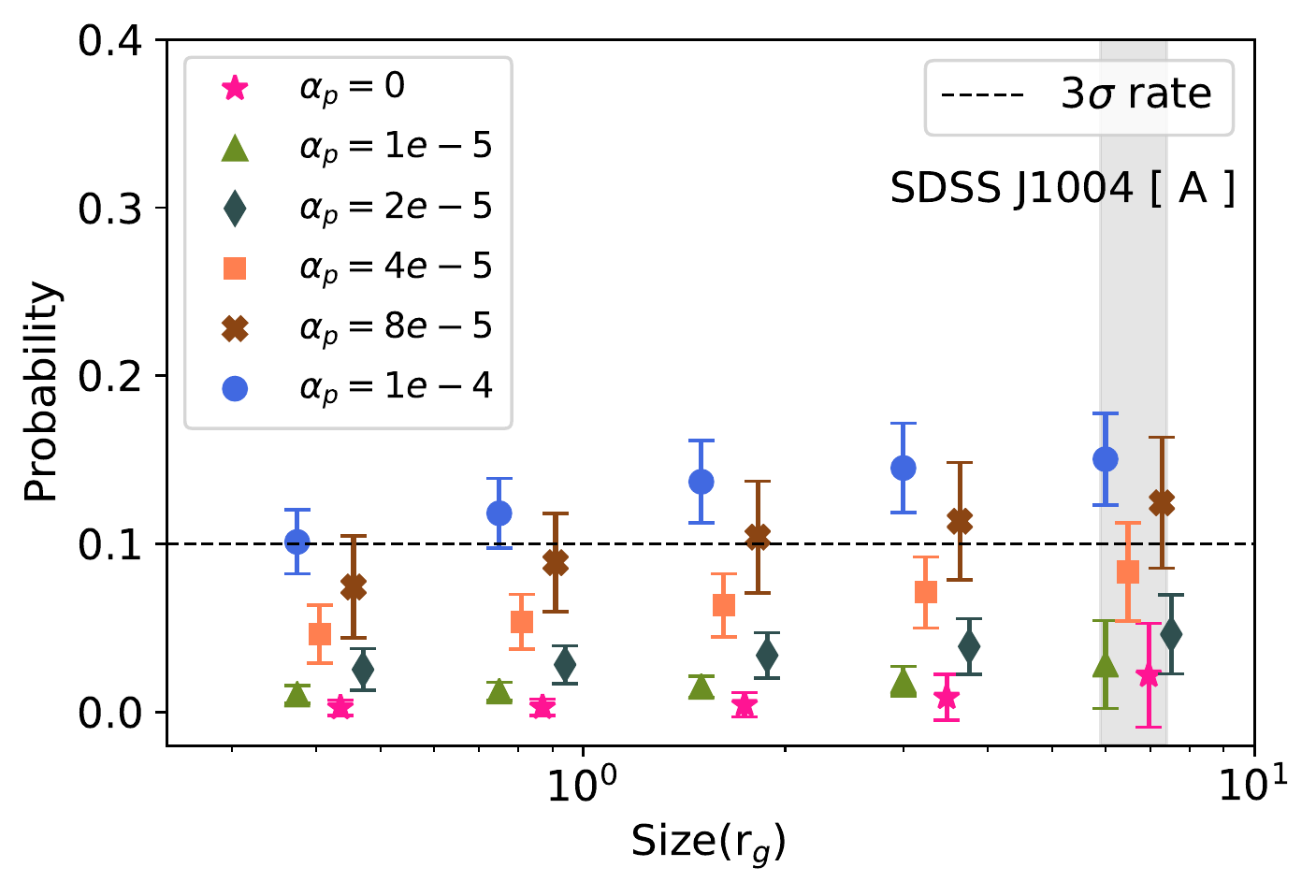}
    \includegraphics[width=0.5\textwidth]{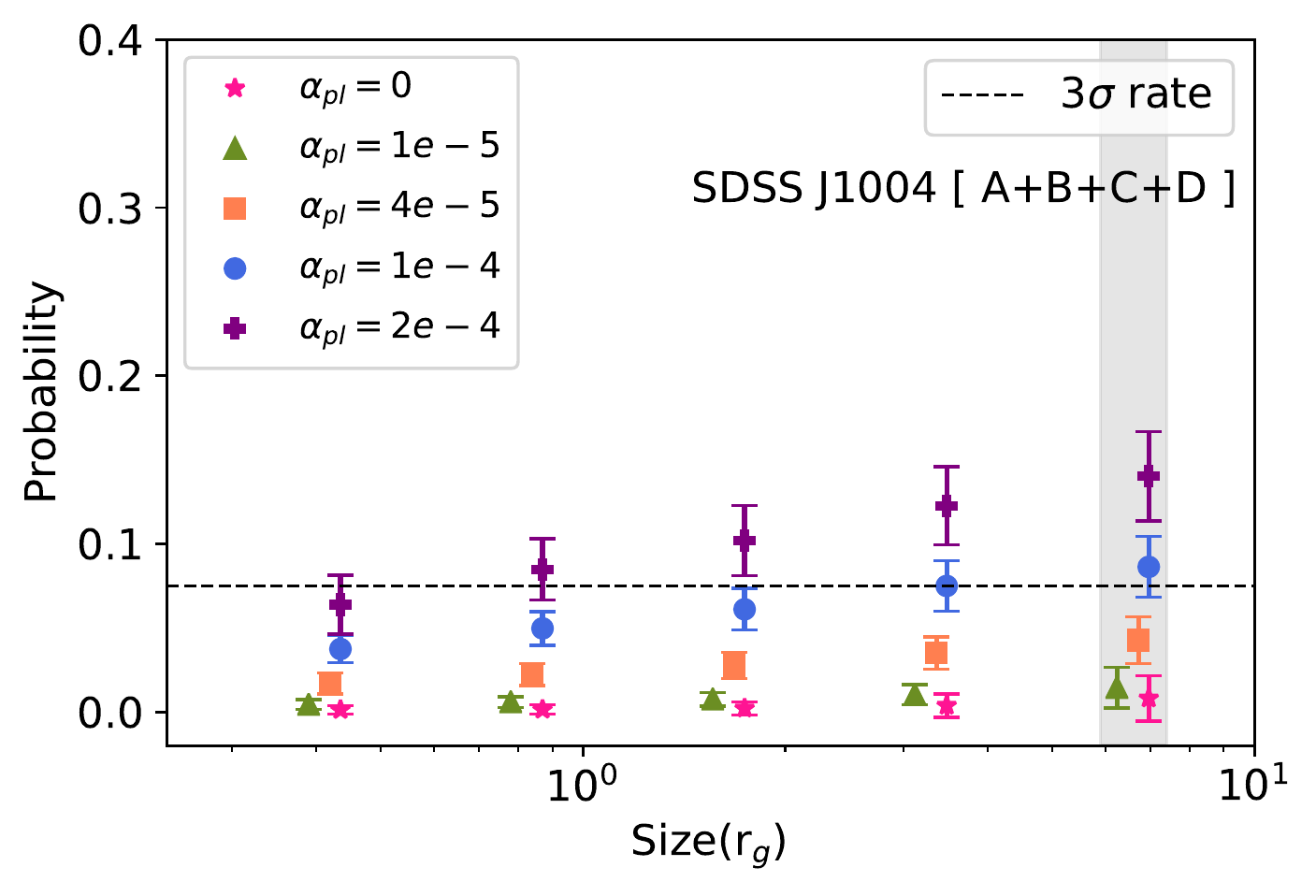}
    \caption{ Model probabilities of observing an \feka\ line energy shift as a function of source size for image A (Left) and combined average of all images (Right) of \qj\ and \sdss\, respectively. Here, the different symbols represent models with the different planet mass fractions ($\alpha_{p}$). The black dashed line mark the observed $>99\%$ detected line shift rates at 3$\sigma$ level. The gray shaded region depicts the recent constraints on the size of the X-ray reflection region by \citet{dai19}. For image A, the stars-only model is ruled out for all source sizes of \qj, and for \sdss, the stars-only model is only viable at the largest source size considered with about $2\sigma$ deviation from the observed rate. For the combined case, the stars-only model is significantly excluded for all source sizes of \qj\ and \sdss. We constrain the halo mass fractions of planets in the mass range of $10^{-8}$--$10^{-3}$~\msun\ to be $10^{-4} < \alpha_{p} < 6\times10^{-4}$ and $4\times10^{-5} < \alpha_{p} < 2\times10^{-4}$ for \qj\ and \sdss, respectively.
    \label{image4}}
\end{figure}

Figure~\ref{image4} shows the model caustic encounter probability from our microlensing analysis for different planet mass fractions and \feka\ source sizes.  The model probability shows an increase with additional planetary population and the source size for both \qj\ and \sdss.  Hereafter, we have conservatively used the largest source size to test the viability of the stars-only model and constrain the planet fraction, $\alpha_{p}$.  This source size is consistent with the recent constraints on the X-ray reflection region of between 5.9--7.4 $r_g$ \citep{dai19}.
We compare the model predictions with the observed line shift rates from the $>99\%$ detected and $>3\sigma$ shifted lines for the highest S/N  image A and the combined average of all images. 
Based on the predictions for image A of  \qj, we find that the stars-only model can be completely ruled out by 16.6 $\sigma$, as it fails to explain the observed line shift rate. As for the image A of \sdss, the stars-only model can only be marginally ruled out as it is at 2.6 $\sigma$ below the observed line shift rate. The analysis results from the combined images disfavor the exclusively stellar scenario at the significance levels of 13.284 $\sigma$ and 4.95 $\sigma$ for \qj\ and \sdss, respectively. For models with additional planets, the model predictions at a reference size of 6 $r_g$ show an ascent with increasing $\alpha_{p}$ until a model is congruent with the observed event rate. In case of \qj, we constrain the planet mass fraction with respect to total halo mass in the mass range of $10^{-8}$--$10^{-3}$~\msun\ as $10^{-4} < \alpha_{p} < 6\times10^{-4}$, and for \sdss, the constraint is $2\times10^{-5} < \alpha_{p} < 10^{-4}$.
 
 \begin{figure}
    \begin{center}
    \includegraphics[scale=1]{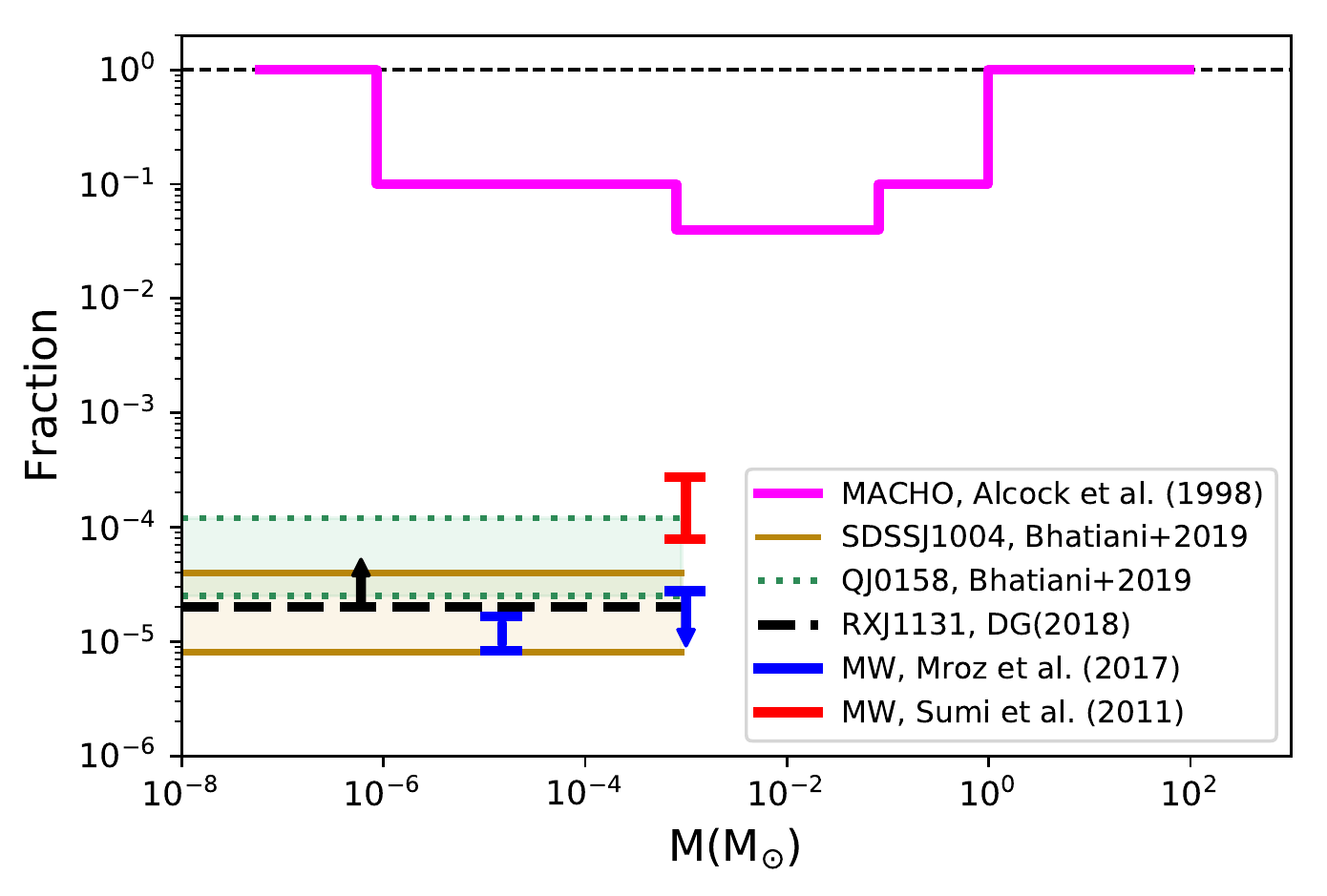}
    \end{center}
    \caption{Mass fraction of unbound planets in the galactic halo is studied as a function of mass in the sub-stellar regime. The constraints presented in this work are shown in gold and green for Q\,J0158$-$4325 and SDSS\,J1004+4112, respectively, with a bin size of one decade in mass. Previous constraints at super earth mass and Jupiter mass for Milky Way by \citet{mroz17} are shown in blue. Constraints for Jovian planets in the MW by \citet{sumi11} are shown in red. Prior constraints on MACHOs by \citet{alcock98} ranging from Moon mass to 100$M_{\odot}$ have been plotted.  We also find that our results are consistent with the constraints from \citet{dai18}, as shown in black, on free-floating planets from Moon to Jupiter mass range for RX\,J1131$-$1231.
    \label{image5}}
\end{figure}

 \section{Discussion} \label{sec:discuss}
 We infer a population of planet-mass objects in two extragalactic systems, one lens galaxy \qj\ at $z=0.317$ and one lens cluster \sdss\ at $z=0.68$, with the leading candidates being free-floating planets and/or primordial black holes.  
We have excluded the stars-only model for \qj\ and \sdss\ and accepted the scenario that favors the combination of stellar and unbound planetary microlenses.
 Together with the previous detection in \rxj\ \citep{dai18} and within the Milky Way \citep{mroz17}, our analysis results suggest that unbound planet-mass compact objects are universal in galaxies.
 
 The bound planets do not contribute to the lensing signal, because at extragalactic lensing scales, their distance to the parent star is too small compared to the Einstein ring of the star.
 The cold dark matter sub-halos potentially can also extend to the planet-mass range, 
 and assuming a halo mass function of $dn/dM \propto M^{-1.8}$ \citep[e.g.,][]{moore99, kly99, helmi02}, 0.05\% of the halo mass will be in the Moon to Jupiter mass range, which is consistent with our constraints.
 However, with an NFW density profile, they will not be efficient micro or nano-lenses to produce the observed lensing signatures.
 
 It is important to distinguish the cross sections for microlensing high magnification events and line shift events.  For high magnification events, the cross section is proportional to the area of Einstein ring, which is proportional to the mass of the microlens.
 For line shift events, we need a caustic on the source region, and thus the cross section is proportional to $m^{1/2}\Delta R$, where $\Delta R$ is the characteristic size of the emission region. Smooth high magnification regions bound by the caustic curves are not conducive to these line shift events. Since we constrain the mass fraction of the planet population to be several orders of magnitude smaller than the stellar population, this implies that most of the high magnification area produced by stars are in the smoothly varying regions of the magnification map, using the ultra-compact \feka\ emission region as a size reference.
 This can be seen in the analytic values of lens number, total mass, and caustic density from Figure~\ref{image2}.
 This leads to an important, testable prediction for this model --- for the \feka\ line shifts produced by the relatively isolated microlensing caustics produced by planets, the optical or even the X-ray continuum flux will experience small to moderate microlensing flux magnification, because the optical continuum emission size is much larger and not sensitive to the magnification produced by planet-mass lenses. 
 I.e., the \feka\ line shifts accompanied by little to moderate flux magnification in the optical or X-ray bands will be the further confirmation for the presence of planet-mass objects.
 This does not mean that those line shifts occurring with large optical or X-ray continuum flux magnification are not caused by planet-mass objects, because the caustics produced by stars and planets can be clustered together.
 Additionally, since the \chandra\ observations are semi-randomly scheduled, it is important to compare the line shift rate with the model caustic density detailed in this paper.  An alternative method counts the number of caustics encountered along a track in the magnification pattern corresponding to the monitoring length and compare with the observed number of line shifts.  This method assumes that the observations occurred exactly at the caustic-crossing time, which is not representing the real situation.  In essence, it is the caustic density rather than number that matters, when the observations are randomly scheduled.
 
Another source of uncertainty is related to the detection threshold of the shifted \feka\ lines.  Here, we include only lines that are detected at the greater than 99\% confidence limit, when calculating the line shift rates.
 This excludes a significant fraction of the shifted lines detected at lower significance in 90--99\% range.  Although each of these individual lines are less reliable, collectively, the presence of these weaker lines are quite convincing.  Including these weaker lines will change the line shift rates and the constraints on the planet fractions. 
In addition, our analysis is based on either the line shift rates in the brightest image or the average of all images.  While focusing on the brightest image subjects less on the line detection bias but more on microlensing biases, because the brightest image tends to be more microlensing active, using the average rate from all images subjects more to the detection biases from those fainter images.
 Therefore, new and deeper observations are still needed to better characterize those weaker lines and measure the line shift rates.
 We summarize the current constraints on unbound planet mass object in galaxies and ICL of a cluster in Figure~\ref{image5}, including those from this paper for \qj\ and \sdss, \rxj\ of \citet{dai18}, and Milky Way measurements at super Earth and Jupiter regime \citep{sumi11, mroz17}.  
 Here, we divide the planet mass fraction to each decade of mass interval to facilitate an easy comparison with other studies.
 Although these initial measurements of unbound planet mass fraction can subject to systematic uncertainties and can be improved in future analysis with better data and modeling, these measurements converge at a planet mass fraction of $\sim 10^{-5}$ with respect to the total mass of the galaxy per decade of mass interval.
 This should be compared with future, more precise theoretical predictions of planet formation and scattering models and determine whether FFPs are sufficient to explain the measured unbound planet mass fractions.
 Our measured planet mass fraction of $\sim 10^{-5}$ per decade mass can also serve as an upper limit for more exotic objects such as primordial black holes, since we expect FFPs will contribute significantly to the unbound population. This is the most stringent limit for primordial black hole demographics at the mass range of $10^{-8}$--$10^{-3}\msun$.

\acknowledgements
We thank S.~Mao and S.~Dong for stimulating discussion, C.~W.~Morgan for providing the macro lens parameters for \qj, and the anonymous referee for helpful comments.
We acknowledge the financial support from the NASA ADAP program NNX17AF26G, NSF grant AST-1413056, and SAO grants AR7-18007X and GO7-18102B.
The microlensing magnification maps were computed at the OU Supercomputing Center for Education \& Research (OSCER) at the University of Oklahoma (OU).

%% This command is needed to show the entire author+affilation list when
%% the collaboration and author truncation commands are used.  It has to
%% go at the end of the manuscript.
%\allauthors

%% Include this line if you are using the \added, \replaced, \deleted
%% commands to see a summary list of all changes at the end of the article.
%\listofchanges


\begin{thebibliography}{}
%A
\bibitem[Alcock et al.(1998)]{alcock98}Alcock, C., et al. 1998, \apj, 499, L9
\bibitem[Anguita et al.(2008)]{ang8}Anguita, T., Schmidt, R. W., Turner, E. L., et al. 2008, A\&A, 480, 327
%B
\bibitem[Bate et al.(2011)]{bate11} Bate, N. F., Floyd, D. J. E., Webster, R. L., \& Wyithe, J. S. B. 2011, \apj, 731, 71
\bibitem[Beuchert et al.(2017)]{beu17}Beuchert, T., Markowitz, A. G., Dauser, T., et al. 2017, A\&A, 603, A50
\bibitem[Blackburne et al.(2014)]{black14} Blackburne, J.~A., Kochanek, C.~S., Chen, B., Dai, X., \& Chartas, G.\ 2014, \apj, 789, 125
%\bibitem[Blackburne et al.(2011)]{black11} Blackburne, J. A., Pooley, D., Rappaport, S., \& Schechter, P. L. 2011b, \apj, 729, 34
\bibitem[Brenneman \& Reynolds(2006)]{br6} Brenneman, L.~W., \& Reynolds, C.~S.\ 2006, \apj, 652, 1028
%C
%\bibitem[Cackett et al.(2014)]{cac14}Cackett, E. M., Zoghbi, A., Reynolds, C., Fabian, A. C., Kara, E., Uttley, P., Wilkins, D. R. 2014, \mnras, 438, 2980
\bibitem[Chang \& Refsdal(1979)]{cr79}Chang, K., \& Refsdal, S. 1979, \nat, 282, 561
%\bibitem[Chang \& Refsdal(1984)]{cr84}Chang, K. & Refsdal, S. 1984, A&A, 132, 168
\bibitem[Chartas et al.(2009)]{char9} Chartas, G., Kochanek, C.~S., Dai, X., Poindexter, S., \& Garmire, G.\ 2009, \apj, 693, 174
\bibitem[Chartas et al.(2012)]{char12} Chartas, G., Kochanek, C.~S., Dai, X., et al.\ 2012, \apj, 757, 137
\bibitem[Chartas et al.(2017)]{char17} Chartas, G., Krawczynski, H., Zalesky, L., et al.\ 2017, \apj, 837, 26
\bibitem[Chen et al.(2011)]{chen11} Chen, B., Dai, X., Kochanek, C.~S., et al.\ 2011, \apjl, 740, L34
\bibitem[Chen et al.(2012)]{chen12} Chen, B., Dai, X., Kochanek, C.~S., et al.\ 2012, \apj, 755, 24
%\bibitem[Chiang et al.(2015)]{chiang15}Chiang, C.-Y., Walton, D. J., Fabian, A. C., Wilkins, D. R., & Gallo, L. C. 2015, /mnras, 446, 759
%D
\bibitem[Dai et al.(2010)]{dai10} Dai, X., Kochanek, C.~S., Chartas, G., et al.\ 2010, \apj, 709, 278
\bibitem[Dai \& Guerras(2018)]{dai18} Dai, X. \& Guerras, E. 2018, \apj, 853, L27
\bibitem[Dai et al.(2019)]{dai19} Dai, X., Steele, S., Guerras, E., Morgan, C.~W., \& Chen, B.\ 2019, arXiv:1901.06007 
\bibitem[Dewangen et al.(2003)]{dew3} Dewangan, G. C., Griffiths, R. E., \& Schurch, N. J. 2003, ApJ, 592, 52
%F
%\bibitem[Fabian et al.(2009)]{fb9}Fabian, A. C., Zoghbi, A., Ross, R. R., et al. 2009, \nat, 459, 540
%\bibitem[Fabian et al.(2012)]{fb12}
\bibitem[Fabian et al.(1995)]{fb95} Fabian, A. C., Nandra, K., Reynolds, C. S., Brandt, W. N., Otani, C., Tanaka, Y., Inoue, H., \& Iwasawa, K. 1995, MNRAS, 277, L11
%\bibitem[Fabian et al.(2006)]{fb6} Fabian, A. C. 2006, \AN, 327, 943
%\bibitem[Fabian \& Vaughan(2003)]{fv03} Fabian, A.~C., \& Vaughan, S.\ 2003, \mnras, 340, L28
\bibitem[Faure et al.(2009)]{faure9} Faure, C., Anguita, T., Eigenbrod, A., et al. 2009, A\&A, 496, 361 
%G-H-I-J
%\bibitem[Gaudi(2012)]{gaudi12} Gaudi, B.~S.\ 2012, \araa, 50, 411
%\bibitem[Gould \& Loeb(1992)]{gl92} Gould, A., \& Loeb, A.\ 1992, \apj, 396, 104
\bibitem[Guerras et al.(2013)]{gue13}Guerras, E., Mediavilla, E., Jimenez-Vicente, J., et al. 2013, ApJ, 764, 160 
\bibitem[Guerras et al.(2017)]{gue17} Guerras, E., Dai, X., Steele, S., et al.\ 2017, \apj, 836, 206
\bibitem[Guerras et al.(2018)]{gue18} Guerras, E., Dai, X., \& Mediavilla, E.\ 2018, arXiv:1805.11498 
\bibitem[Helmi et al.(2002)]{helmi02} Helmi, A., White, S.~D., \& Springel, V.\ 2002, \prd, 66, 063502
\bibitem[Inada et al.(2003)]{inada3} Inada, N., et al. 2003, Nature, 426, 810
\bibitem[ Inada et al.(2008)]{inada8}Inada, N., Oguri, M., Falco, E. E., Broadhurst, T. J., Ofek, E. O., Kochanek, C. S., Sharon, K., \& Smith, G. P. 2008, PASJ, 60, L
\bibitem[Ingrosso et al.(2009)]{ingrosso09} Ingrosso, G., Novati, S.~C., de Paolis, F., et al.\ 2009, \mnras, 399, 219 
%\bibitem[Jovanovic et al.(2009)]{jov9} Jovanovic, P., Popovi ´ c, L. ´ C., & Simi ˇ c, S. 2009, ´ New Astron. Rev., 53, 156
%K
\bibitem[Kara et al.(2014)]{kara14}Kara, E., Fabian, A. C., Marinucci, A., et al. 2014, MNRAS, 445, 56
%\bibitem[Kara et al.(2016)]{kara16}Kara, E., Alston, W. N., Fabian, A. C., et al. 2016, MNRAS, 462, 511
%\bibitem[Keeton et al.(2001)]{keet1}Keeton, C. R. 2001a, preprint (astro-ph/010234)
 \bibitem[Kayser et al.(1986)]{key86}Kayser, R., Refsdal, S., Stabell R. 1986, A\&A, 166, 36
\bibitem[Klypin et al.(1999)]{kly99} Klypin, A., Kravtsov, A.~V., Valenzuela, O., \& Prada, F.\ 1999, \apj, 522, 82
\bibitem[Kochanek(2004)]{koch4} Kochanek, C.~S.\ 2004, \apj, 605, 58
%\bibitem[Kochanek et al.(2006)]{ko6}
%\bibitem[Kochanek et al.(2007)]{koch7} Kochanek, C.~S., Dai, X., Morgan, C., Morgan, N., \& Poindexter, S.~C., G.\ 2007, Statistical Challenges in Modern Astronomy IV, 371, 43
\bibitem[Krawczynski \& Chartas(2017)]{kc17} Krawczynski, H., \& Chartas, G.\ 2017, \apj, 843, 118
%L-M-N
\bibitem[Ledvina et al.(2018)]{led18} Ledvina, L., Heyrovsk{\'y}, D., Dovčiak, M. 2018, \apj, 863, 66
%\bibitem[Lewis et al.(1998)]{lew98}
\bibitem[Luhman(2012)]{luhman12} Luhman, K. L. 2012, ARA\&A, 50, 65
%\bibitem[Mao \& Paczynski(1991)]{mp91} Mao, S., \& Paczynski, B.\ 1991, \apjl, 374, L37
%\bibitem[MacLeod et al.(2015)]{mac15}MacLeod, C. L., Morgan, C. W., Mosquera, A., et al. 2015, \apj, 806, 258
\bibitem[Maza et al.(1995)]{maza95} Maza, J., Wischnjewsky, M., Antezana, R., \& González, L. E. 1995, RMxAA, 31, 119
\bibitem[Mediavilla et al.(2006)]{med6}Mediavilla, E., Muñoz, J. A., Lopez, P., et al. 2006, ApJ, 653, 942
\bibitem[Mediavilla et al.(2011)]{med11} Mediavilla, E., Mediavilla, T., Mu{\~n}oz, J.~A., et al.\ 2011, \apj, 741, 42
\bibitem[Mediavilla et al.(2011b)]{med11b}Mediavilla, E., Munoz, J. A., Kochanek, C. S., et al. 2011b, ApJ, 730, 16
\bibitem[Moore et al.(1999)]{moore99} Moore, B., Ghigna, S., Governato, F., et al.\ 1999, \apjl, 524, L19
\bibitem[Morgan et al.(1999)]{morg99} Morgan, N. D., Dressler, A., Maza, J., Schechter, P. L., \& Winn, J. N. 1999, AJ, 118, 1444
\bibitem[Morgan et al.(2008)]{morg8} Morgan, C.~W., Kochanek, C.~S., Dai, X., Morgan, N.~D., \& Falco, E.~E.\ 2008, \apj, 689, 755-761
\bibitem[Morgan et al.(2012)]{morg12}Morgan, C. W., Hainline, L. J., Chen, B., et al. 2012, ApJ, 756, 52
%\bibitem[Mosquera et al.(2013)]{mosq13} Mosquera, A.~M., Kochanek, C.~S., Chen, B., et al.\ 2013, \apj, 769, 53
\bibitem[Mr{\'o}z et al.(2017)]{mroz17} Mr{\'o}z, P., Udalski, A., Skowron, J., et al.\ 2017, \nat, 548, 183
\bibitem[Mr{\'o}z \& Poleski(2018)]{mp18} Mr{\'o}z, P., \& Poleski, R.\ 2018, \aj, 155, 154 
\bibitem[Niikura et al.(2019)]{niikura19} Niikura, H., Takada, M., Yasuda, N., et al.\ 2019, Nature Astronomy
%O-P
\bibitem[O{'}Dowd et al.(2015)]{odow15}O’Dowd M. J., Bate N. F., Webster R. L., Labrie K., Rogers J., 2015, ApJ, 813, 6
\bibitem[Oguri(2004)]{og4}Oguri, M., Inada, N., Keeton, C. R., et al. 2004, ApJ, 605, 78
\bibitem[Oguri(2010)]{og10} Oguri, M. 2010, PASJ, 62, 1017
\bibitem[Oguri et al.(2018)]{og18} Oguri, M., Diego, J. M., Kaiser, N., Kelly, P. L., \& Broadhurst, T. 2018, PhRvD, 97, 023518
\bibitem[Ota et al.(2006)]{ota6} Ota, N., et al. 2006, ApJ, 647, 215
\bibitem[Paczy{\'n}ski(1986)]{pac86}Paczynski, B. 1986, ApJ, 301, 503
%\bibitem[Parker et al.(2014)]{park14}Parker, M. L., Wilkins, D. R., Fabian, A. C., et al. 2014, MNRAS, 443, 1723
\bibitem[Popovi{\'c} et al.(2003)]{pop3}Popović, L. Č., Mediavilla, E. G., Jovanović, P., \& Muñoz, J. A. 2003b, A\&A, 398, 975
%\bibitem[Popovi{\'c} et al.(2005)]{pop5}Popopvi´c, L. C, Jovanovi´c, Mediavilla, E., Zakharov, A. F., ˇ Abajas, C., Mu˜noz, J. A., Chartas, G. 2006, ApJ, 637, 620
\bibitem[Popovi{\'c} et al.(2006)]{pop6} Popovi{\'c}, L.~{\v C}., Jovanovi{\'c}, P., Mediavilla, E., et al.\ 2006, \apj, 637, 620
\bibitem[Pooley et al.(2007)]{poo7}Pooley, D., Blackburne, J. A., Rappaport, S., \& Schechter, P. L. 2007, ApJ, 661, 19
\bibitem[Pounds et al.(2003)]{pound3}Pounds K.A., Reeves J.N., Page K.L., Wynn G.A., O’Brien P.T., 2002, MNRAS, 342, 1147

%Q-R-S-T-U
\bibitem[Rasio et al.(1996)]{rasio96} Rasio, F. A. \& Ford, E. B. 1996, Science, 274, 954
\bibitem[Reynolds et al.(2000)]{rey20}Reynolds, C. S., \& Wilms, J. 2000, ApJ, 533, 821
%\bibitem[Schechter \& Wambsganss(2002)]{sch2} Schechter, P. L., & Wambsganss, J. 2002, \ApJ, 580, 685
\bibitem[Sharon et al.(2005)]{sharon5}Sharon, K., et al. 2005, ApJ, 629, L73
%\bibitem[Sluse et al.(2003)]{sluse03} Sluse, D., Surdej, J., Claeskens, J.-F., et al.\ 2003, \aap, 406, L43
\bibitem[Sluse et al.(2012)]{sluse12}Sluse, D., Hutsemekers, D., Courbin, F., Meylan, G., \& Wambsganss, J. 2012, A\&A, 544, A62
\bibitem[Strigari et al.(2012)]{strigari12} Strigari, L.~E., Barnab{\`e}, M., Marshall, P.~J., \& Blandford, R.~D.\ 2012, \mnras, 423, 1856
\bibitem[Sumi et al.(2011)]{sumi11} Sumi, T., Kamiya, K., Bennett, D.~P., et al.\ 2011, \nat, 473, 349
\bibitem[Turner et al.(2002)]{turn2}Turner, T. J., et al. 2002, ApJ, 574, L123
%\bibitem[Uttley et al.(2014)]{utt14} Uttley, P., Cackett, E. M., Fabian, A. C., Kara, E., Wilkins, D. R. 2014, A&ARv, 22, 72

%V-W-X-Y-Z
\bibitem[Vaughan \& Fabian(2004)]{vaug4} Vaughan, S., \& Fabian, A. C. 2004, MNRAS, 348, 1415
\bibitem[Venumadhav et al.(2017)]{venu17} Venumadhav, T., Dai, L., \& Miralda-Escudé, J. 2017, ApJ, 850, 49
\bibitem[Veras \& Raymond(2012)]{verray12} Veras, D., \& Raymond, S. N. 2012, MNRAS, 421, L117
\bibitem[Wambsganss et al.(1990a)]{wam90a} Wambsganss, J. 1990, in Lecture Notes in Physics, Berlin Springer Verlag, Vol. 360, Gravitational Lensing, ed. Y. Mellier, B. Fort, \& G. Soucail, 186–191 Wambsganss, J., Witt, H. J., \& Schneider, P. 1992, A\&A, 258, 591
%\bibitem[Wambsganss et al.(1992)]{wam92} Wambsganss, J. 1992, ApJ, 386, 19
\bibitem[Wambsganss(2001)]{Wam1} Wambsganss, J. 2001a, in Microlensing 2000: A new Era of Microlensing Astrophysics, ed. J.W.Menzies and P.D.Sackett ASP Conf. Series, 239, 351
%\bibitem[Wambsganss(2006)]{wam6} Wambsganss, J.\ 2006, Saas-Fee Advanced Course 33: Gravitational Lensing: Strong, Weak and Micro, 453
\bibitem[Weidenschilling \& Marzari(1996)]{weidzari96} Weidenschilling, S. J. \& Marzari, F. 1996, Nature, 384, 619
\bibitem[Witt(1990)]{witt90} Witt, H.~J.\ 1990, \aap, 236, 311 
%\bibitem[Zoghbi et al.(2010)]{zoghbi10} Zoghbi, A., Fabian, A. C., Uttley, P., et al. 2010, MNRAS, 401, 2419 Zoghbi A., Fabian A. C., Reynolds C. S., Cackett E. M. 2014, MNRAS, 422, 129
%\bibitem[Zoghbi et al.(2014)]{zoghbi14} Zoghbi A., Fabian A. C., Reynolds C. S., Cackett E. M. 2014, MNRAS, 422, 129
\bibitem[Young et al.(2005)]{young5} Young, A. J., Lee, J. C., Fabian, A. C., Reynolds, C. S., Gibson, R. R., \& Canizares, C. R. 2005, ApJ, 631, 733
\end{thebibliography}
\end{document}